\documentclass[11pt]{article}
\usepackage[utf8]{inputenc}
\usepackage{graphicx}
\usepackage{subfig}
\usepackage{amsmath}
\usepackage{geometry}
\usepackage{bbm}
\usepackage{multicol}
\usepackage{natbib}
\usepackage{tabularx}
\usepackage{enumitem}
\usepackage{array}
\usepackage{multirow}
\usepackage{parskip}
\usepackage{hyperref}

\providecommand{\keywords}[1]
{
  \small	
  \textbf{\textit{Keywords---}} #1
}

\title{Investigating the `old boy network' using latent space models}

\author{Ian Hamilton, University of Warwick, Coventry, U.K.}
\date{}

\begin{document}

\maketitle







\section*{Abstract}
This paper investigates the nature of institutional ties between a group of English schools, including a large proportion of private schools that might be thought of as contributing to the `old boy network'. The analysis is based on a network of bilaterally-determined school rugby union fixtures. The primary importance of geographical proximity in the determination of these fixtures supplies a spatial `ground truth' against which the performance of models is assessed. A Bayesian fitting of the latent position cluster model is found to provide the best fit of the models examined. This is used to demonstrate a variety of methods that together provide a consistent and nuanced interpretation of the factors influencing community and edge formation in the network. The influence of homophily in fees and the proportion of boarders is identified as notable, with evidence that this is driven by a community of schools having the highest proportion of boarders and charging the highest fees, suggestive of the existence and nature of an `old boy network' at an institutional level. 

\keywords{clustering; community detection; latent position cluster model; relative importance; schools; sport}

\section{Introduction}

`Old boy network' is an English phrase used to refer to the informal system through which men assist other men of a similar socio-economically privileged background, reflected in attending the same school or university. It derives from the term, `old boy', used to refer to a former pupil at a British `public school', the confusing name given to a subset of traditional, and often especially expensive, private schools in England. Generally the old boy network is considered to act at an individual level, but it may be reasonable to think that it will be stronger if and where there are institutional links between schools. However, it is not clear to what degree there are such links and, to the degree that there are, what features of the school might define them. In the English school system, there are a number of distinctions that might be thought to be identifiable in the functioning of such a network, for example private/state, boarding/day, level of fees. In this paper, latent space models are applied to a network of school rugby union fixtures to investigate whether such functional networks exist and, if they do, what the nature of them may be.

The fixture data provides an interesting data set on which to apply these approaches, for a couple of reasons. First, since one may reasonably expect that a primary consideration in agreeing a fixture would be geographical proximity, then there is a spatial `ground truth' against which to assess the models considered. Second, there is reason to believe that the set of fixtures may be informative. Fixtures change from year to year but not drastically, and the process of fixtures being scheduled and changed has been taking place for many decades, with the first school rugby union fixture taking place over a hundred years ago. Together these considerations suggest that there has been time for relevant factors to exert an influence such that the observed situation represents a steady state with respect to the schools' current relationships, allowing inferences to be informative. 

Network latent space models position nodes in an unobserved latent space, with the probability of an edge existing between any two nodes being related to the proximity of the two nodes in the latent space. Modelling in this way can allow a number of features common to networks to be captured --- transitivity; homophily by attributes; and clustering --- as well as often allowing for informative graphical representations. 


In the context of networks, transitivity is the phenomenon that two nodes, which each share an edge with the same third node, will have a higher probability of having an edge with each other than a pair that do not. Homophily by attributes describes the greater propensity for an edge to exist when two nodes share observed attributes. For example in a friendship network these could be attributes such as age, sex, geographical location, or recreational interests. Transitivity and homophily by attributes will both lead to clustering, but it is not uncommon to observe clustering beyond what may be explained by these features. This may be due to homophily by unobserved attributes, self-organisation of actors, the popularity of particular actors, or endogenous attributes such as position in the network \citep{handcock2007model}.

Early latent space network models used multidimensional scaling, and while these captured transitivity and homophily by attributes they were reliant on the arbitrary choice of a distance measure, leading to variable interpretations. \citet{hoff2002latent} proposed a stochastic model in which the latent space positions may be estimated through standard statistical techniques. \citet{handcock2007model} extended that model to account for clustering by assuming that the latent positions are drawn from a finite mixture of multivariate normal distributions. 

In the present setting there is reason to believe that there will be clustering beyond that due to transitivity or observed homophily of attributes. First, because there are likely to be factors not captured in available data that account for unobserved homophilies, such as sport orientation within school culture, and social networks between relevant staff. Second, because the fixture information is publicly accessible, which might increase the potential for further clustering through self-organisation; if a school sees that a number of the schools they play have an opponent in common who they do not play they may consider proposing a fixture.

The paper proceeds in Section \ref{sec:Data} by discussing the data, with details of its collection, and a brief analysis highlighting some features and inter-relations of the school covariates considered. In Section \ref{LS model} the latent space models of \cite{hoff2002latent} and \cite{handcock2007model} are fitted. In Section \ref{sec:LS Covariates} the influence of the different covariates is explored. Section \ref{sec:Concluding Remarks} provides some concluding remarks. 

\section{Data} \label{sec:Data}

\subsection{Background} \label{sec: Data background}
The Daily Mail Trophy is an annual tournament between some of the best school rugby teams in England, along with a single school in Wales. In order to qualify for a ranking in the Daily Mail Trophy a school must register for the tournament and compete against a minimum of five other teams in the tournament. There are in total 118 schools included, playing between 3 and 37 matches over the course of the three seasons. The matches played as part of the tournament are typically only a subset of the matches played by these schools, with other matches taking place as friendlies with non-tournament schools, or as part of a centrally scheduled knock-out competition. In almost all cases however, if they have played a match against one of the other tournament teams, outside of the centrally scheduled knock-out competition, it would be classed as a Daily Mail Trophy match and would appear in the data set. So the existence or absence of a match with another school is an accurate representation of the bilaterally arranged fixtures within the set of schools in the tournament. 

The network to be investigated is based on all matches in the Daily Mail Trophy in the 2015--16, 2016--17 and 2017--18 seasons. Schools who registered but were unable to complete five eligible matches in any given season are maintained in our analysis, despite being excluded from an official ranking. The network being considered is undirected; each node is a school, and each edge has a value equal to the number of seasons, out of the three for which there are data, in which the two schools played each other. This network is considered in relation to school-level data representing variables that could contribute to some observable homophily.

\subsection{Data collection}
The fixtures and results for the Daily Mail Trophy were kindly shared by Schoolsrugby.co.uk, the organiser of the tournament, but are also available at the tournament website (\url{www.schoolsrugby.co.uk/dailymailtrophy.aspx}). In the analysis that follows it is the number of seasons, out of the three considered, that two teams play each other that are used, removing the five instances where teams played each other twice in a season.

For each school, data were also collected on: 
\begin{enumerate} \itemsep0em
    \item annual fees (Fees)
    \item year of foundation (Founded)
    \item the number of boys in sixth form (6th Form boys)
    \item the proportion of pupils in the school that are boys (Percent boys)
    \item the proportion of pupils that are boarders (Percent boarders)
    \item whether the school is privately or state funded (School type)
    \item whether the school played one or two terms of rugby (Term type)
    \item performance rating of the rugby-playing strength of the school (Rating)
\end{enumerate}

While it would be plausible to consider other variables, these were either not readily available, for example the size of sports bursaries or proportion of pupils from overseas, or were substantially accounted for by this set of covariates, for example if a school is co-educational throughout or just in sixth form, or not at all, is well captured in the proportion of pupils that are boys and the number of sixth form boys. All covariate data were collected during the first week of June 2019. This means they are not contemporaneous with the fixtures occurring in the period 2015--2018. This is not expected to have materially impacted any conclusions because the covariates are not subject to large year on year changes, .

Fees information was sourced from the schools' own websites. Annual fees were taken to be the minimum fees for full-time education of a pupil in Upper Sixth form, not accounting for bursaries or scholarships. Thus it is zero for state-funded schools, standard day fees for schools that admit day pupils, and minimum boarding fees for schools that admit only boarders. 

Year of foundation was generally more readily available on the Wikipedia page for the school than the school's website and so the Wikipedia date was used. This was originally included as it was suspected that it might define a meaningful similarity between schools. But in collecting the data it became clear that the trajectories of schools were very diverse and it was often even difficult to be sure of a definitive foundation date with occurrences of, for example, schools moving geographically, amalgamating, moving from private to state or vice versa, and renaming not uncommon. All of these contributed to subjectivity in the definition of year of foundation. It is included in the analysis however as it provides a useful sanity check to some of the later methods in the degree to which those methods identify year of foundation as a non-informative covariate.

The data on the number of pupils, including the proportions of boys and boarders, and the absolute numbers of sixth form boys, were generally not available from schools' websites. For privately funded schools, these data were collected from the Independent Schools Council (ISC) website (\url{www.isc.co.uk}). For boys and for girls, this reports the number of boarders, the number of day pupils, and the number of sixth form pupils. From these the proportion of day pupils, and the proportion of boys are calculated. However it should be noted that since schools admit pupils from different ages, the relative proportions as they pertain to a consistent age group, say 13--18, are unlikely to be the same. Given the purpose of including these proportions --- identifying homophilies --- and that no ready alternative was available, then this seems acceptable, as they still provide evidence on the nature of the school. Pupil numbers for state schools were taken based on the most recent relevant government report \citep{Gov2019pupils}. These did not include a delineation for numbers of sixth form students or of boarders. For this, the most recent available Ofsted report with such data was used (\url{www.gov.uk/government/organisations/ofsted}). Since these were from previous years and so total numbers differed, the proportions of sixth form students or boarders were assumed to be constant, and the absolute number of sixth form boys was adjusted for the current total number of pupils. The total number of pupils was never materially different from current numbers, and so one may reasonably be confident that these numbers are accurate. However it should be noted that these Ofsted reports were quite commonly from as much as a decade ago. In one case, Ampleforth College, the population information was not available on the ISC website and so the method used was the same as for the state schools, with the most recent Independent Schools Inspectorate report (\url{www.isi.net}) used, in place of an Ofsted report, to derive the relevant proportions. Whether a school was state or privately funded was determined by its classification in the most recent relevant government report \citep{Gov2019pupils}.

In order to determine if the school played one or two terms of rugby, the month of the school's first and last fixtures of the 2018 season were considered, including those fixtures outside the Daily Mail Trophy competition. If these both lay in a single term, then it was interpreted as being a single term rugby-playing school. If there were two or more matches in the second term then it was deemed to be a two term school. If there was a single match outside of first term then previous seasons' fixtures as well as any information from the school's website was used to make the categorisation.

The performance rating used here is described in \citet{hamilton2021retrodictive}, applied to an aggregation of the three seasons' results. The method accounts for the varying schedule strengths of participating schools in a manner consistent with the predominant league points system used in rugby union. Taking the projected league points per match were each team to play every other team home and away in a round robin format provides a positive-valued measure of a comparable scale to the other covariates as well as an intuitive interpretation to the rating measure.

For all of these factors, when considering nodal covariates the absolute difference is used. The binary variables of school type and term type take value 0 if identical and 1 if different for each pair. The two percentage variables of proportion of boys and proportion of boarders are taken as the absolute difference in these percentages. 

Postcodes were taken from the Daily Mail Trophy website and then used via a Google Maps API in order to calculate travel times and distances between schools using the R package googleway \citep{cooley2017googleway} and to project locations onto a relevant map. Distances were calculated as at Saturday 5th October 2019 12pm using the ``best guess" methodology, assuming a journey by road. In order to plot schools geographically, latitude and longitude for each postcode was sourced from \url{www.freemaptools.com}.

The five continuous variables are scaled to allow for better comparability and interpretability, specifically the following units are used:
\begin{table}[htbp!]
\centering
\begin{tabularx}{\linewidth}{lX}
    Travel Time & hours \\
    Fees & £10,000 \\
    Founded & centuries \\
    6th Form Boys & 100 boys \\
    Rating & projected league points per match 
\end{tabularx}
\end{table}


\subsection{Exploratory Data Analysis}
The two binary covariates of whether a school is privately or state funded, and whether it plays one or two terms of rugby are compared first. Table \ref{tab:P/S vs 1/2} shows that there is a clear relation, with the private schools substantially more likely to play one term and state funded schools more likely to play two.

\begin{table}[hbtp!]
\centering
\begin{tabular}{l|cc}

        & One term & Two terms \\
\hline
Private & 79       & 17        \\
State   & 8        & 14       \\

\end{tabular}
\caption{Number of schools in the tournament playing one or two terms of rugby and private or state funded}
\label{tab:P/S vs 1/2}
\end{table}

\begin{figure}[hbp!]
    \centering
    \includegraphics[width=15cm, height = 15cm]{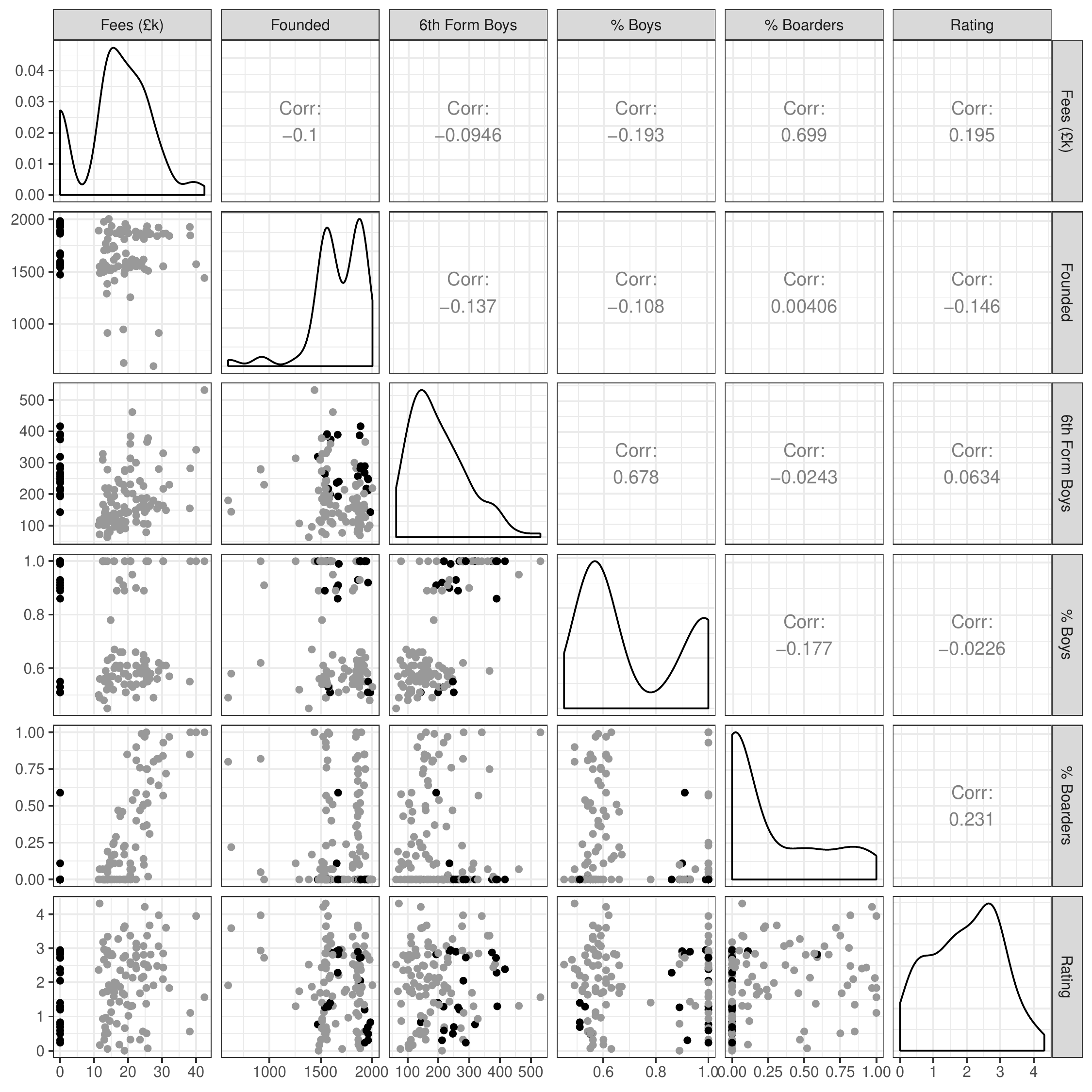}
    \caption{Scatterplots of school covariates. Private schools in grey, state funded schools in black}
    \label{fig:Covariate scatterplots}
\end{figure}

The other covariates are presented in Figure \ref{fig:Covariate scatterplots}. A number of things may be noted. First the individual covariate distributions are largely in line with expectations. A number of the covariates are bimodal. This is unsurprising in the case of fees, with one of the modes at zero, or in the case of the percentage of boys, with modes close to 50\% and 100\%, reflecting a predominance of all-through co-educational and single sex schools, but the bimodal nature to the year of foundation is less intuitive without more historical context. The number of sixth form boys is right-skewed with a mode around 150. The proportion of boarders has a clear mode at just above zero, reflecting a high proportion of day schools and some schools with just a handful of boarders. The remaining proportions are, perhaps not so intuitively, distributed quite evenly up to full boarding status, but without another clearly identifiable mode. Rating is left-skewed with its mode at around 2.7 league points per match, which suggests teams would be expected to share almost one and a half bonus point per match on average. Importantly for our analysis, with the exception of the percentage of boarders and fees, and the proportion of boys and the number of sixth form boys, which have Pearson correlation of 70\% and 68\% respectively, the continuous covariates all have absolute correlations of less than 25\%, which is helpful for being able to discern independent effects. The correlation between the proportion of pupils that are boarders and the fees is not due, as one might first suspect, to boarding fees being higher as they must also account for living expenses, since the minimum upper sixth form fees have been used in all cases, and there are only four pure boarding schools where those minimum fees include boarding fees. In all other cases they represent a day fee. A correlation of 68\% between the proportion of pupils that are boys and the number of sixth form boys is less surprising. Here shading has been used in the scatterplots to differentiate state funded and private schools. Apart from the self-explanatory difference in fees, these also show that within this set of schools the state schools are more likely to have high numbers of sixth form boys, to be single sex (or close to), and to be day schools rather than boarding. We might therefore expect some confounding of these factors in later analyses. These charts were also looked at with a differentiation based on the number of terms of rugby played. This highlighted similar features, as one might expect from Table \ref{tab:P/S vs 1/2}, but less strongly.

\section{Latent Space Model} \label{LS model}
\subsection{Model specification}
Given the three seasons of fixture data, a binomial latent space model is fitted here, the general form of which is
\begin{align*}
    P(A \mid \boldsymbol{Z},\boldsymbol{x},\boldsymbol{\beta}) &= \prod_{i<j}\binom{3}{A_{ij}}\mu_{ij}^{A_{ij}}(1-\mu_{ij})^{3-A_{ij}} \tag{1}\label{eq:1}\\
    \text{logit}(\mu_{ij})
    &= \beta_0 + \sum_{k=1}^{p}x_{ijk}\beta_{k} - d(\boldsymbol{Z_i},\boldsymbol{Z_j}). \tag{2}\label{eq:2}
\end{align*}
Here $A = [A_{ij}]$ is the symmetric adjacency matrix of fixtures with $A_{ij}$ equal to the number of seasons out of the three in which teams $i$ and $j$ played each other, $x_{ijk}$ is the $k$th nodal covariate for teams $i$ and $j$, $\beta_k$ is the coefficient for the $k$th covariate, $\boldsymbol{Z_i}$ is the latent position for team $i$, and $d(\boldsymbol{Z_i},\boldsymbol{Z_j})$ is the Euclidean distance between teams $i$ and $j$ in the latent space. This is the binomial version of the model proposed by \citet{hoff2002latent}.

It is worth noting that a sociality parameter is not included. In this context a sociality parameter would describe the propensity for a particular school to have matches. It is not included here because the fixtures represented are an incomplete set of fixtures for the participating teams, with teams generally playing additional matches, against teams outside of the tournament, as friendlies or as part of other competitions. While these data are not fully available, the schedule for a sample of teams has been inspected and generally they have played a similar total number of matches, so observing a team to have higher degree within the network is not a reflection of that team having a higher propensity to play matches in general. Including such effects could therefore, for example, misleadingly diminish the extent to which one might infer a lower homophily from the absence of a fixture in the case of teams with lower degree. On the other hand, the current rules of the tournament encourage teams to play as many other tournament participants as possible, since they are awarded bonus points merely for playing matches in the tournament, independent of result. Based on conversations with the tournament organisers and observations of historic results, it is likely that there will be a difference in motivation between teams in their desire to be competitive in the tournament, and some may actively seek to schedule more tournament matches. However there are only a small number of teams, consistent across years, for whom the tournament is a goal in itself. For the vast majority, they enter simply because they can, as a by-product of their standard fixture list, which is largely similar from year to year. So while this effect could be argued to be a genuine sociality effect, within the context of just this network of fixtures, it is likely to apply to only a small minority of teams. Therefore on balance, the distorting impact of inclusion is considered to be more of a danger than that of exclusion and so no sociality effect is included. All models are fitted using the latentnet package in R \citep{krivitsky2008fitting}.

\subsection{Hierarchical Clustering}\label{sec:LS Hierarchical}
Initially the parameters of the model represented in equations (1)--(2) are estimated through the method of maximum likelihood and with no covariates included. 

\begin{figure}
    \centering
    \includegraphics[width=13.5cm, height = 9cm]{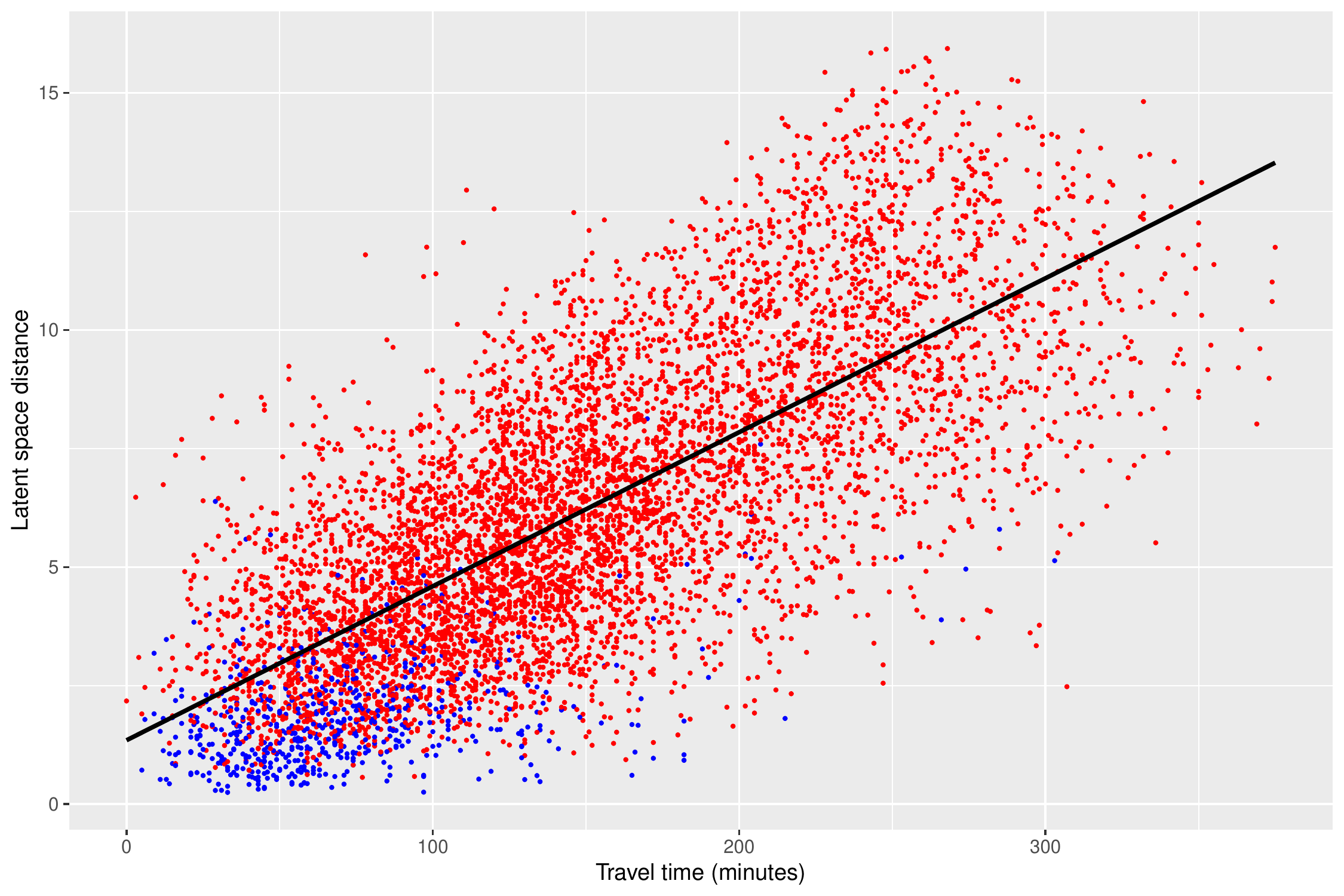}
    \caption{Scatterplot of latent space distance against travel time. School pairs who do not play each other during the three seasons are in red, pairs who play each other at least once are in blue. OLS regression line is shown in black.}
    \label{fig:Model distance vs Travel time}
\end{figure}

It was suspected that geographical proximity would be a primary driver of the propensity for a match to occur, and therefore of model distance. Figure \ref{fig:Model distance vs Travel time} plots the pairwise Euclidean distances, calculated using the latent space positions, against the pairwise estimated travel times. This shows a strong relationship, with a Pearson correlation of 74\%. Travel time is used here as this would seem to be a more relevant motivating condition for a fixture than geographical distance, but substantially similar results are found when using geographical distance. The comparison between latent space distance and travel time may also be used as a means of testing the choice to use a two dimensional latent space. When fitting with three dimensions the correlation increased only by 1\%, and goodness of fit measures based on the posterior predictive distribution of degree and minimal geodesic distance \citep{krivitsky2008fitting} showed no clear improvement with increased dimension, strongly suggesting that modelling in two dimensions, with the representational benefits it brings, is a reasonable choice. As such, all further models will be applied using a two dimensional latent space.

\begin{figure}
    \centering
    \includegraphics[width=13.5cm, height = 9cm]{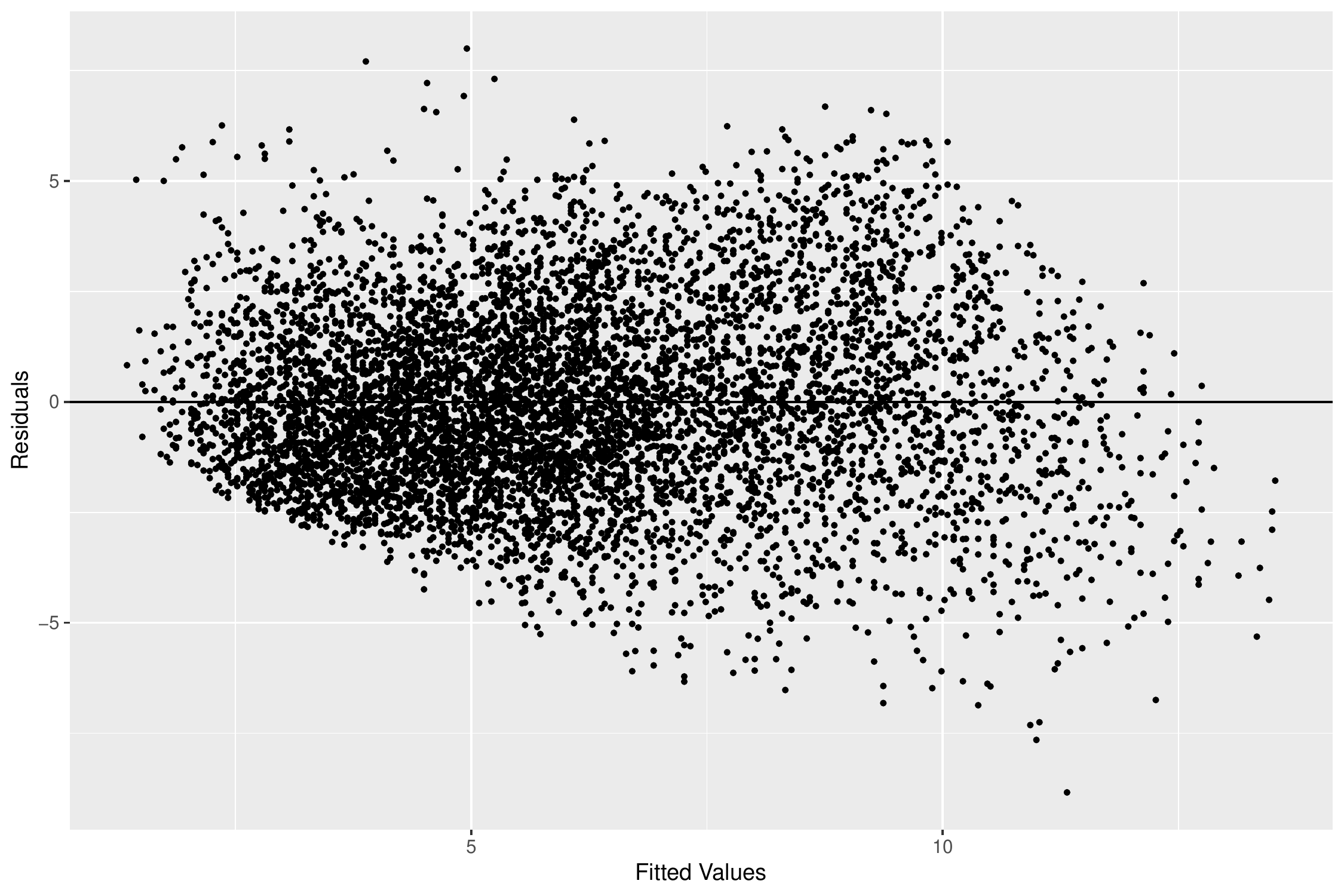}
    \caption{Scatterplot of residuals from OLS fit of latent space distance against travel time}
    \label{fig:Residuals Model distance vs Travel time}
\end{figure}

While the fit is reasonable, there seems to be a notable skew to the residuals. Looking at the residual plot in Figure \ref{fig:Residuals Model distance vs Travel time} it can be seen that this is driven by two groups. The first has high travel time and considerably lower latent space distance than the linear regression would suggest, the second low travel time but considerably higher latent space distance than the regression would suggest. 

\begin{figure}
\centering
	\subfloat{\includegraphics[width=6.5cm,height=6.5cm]{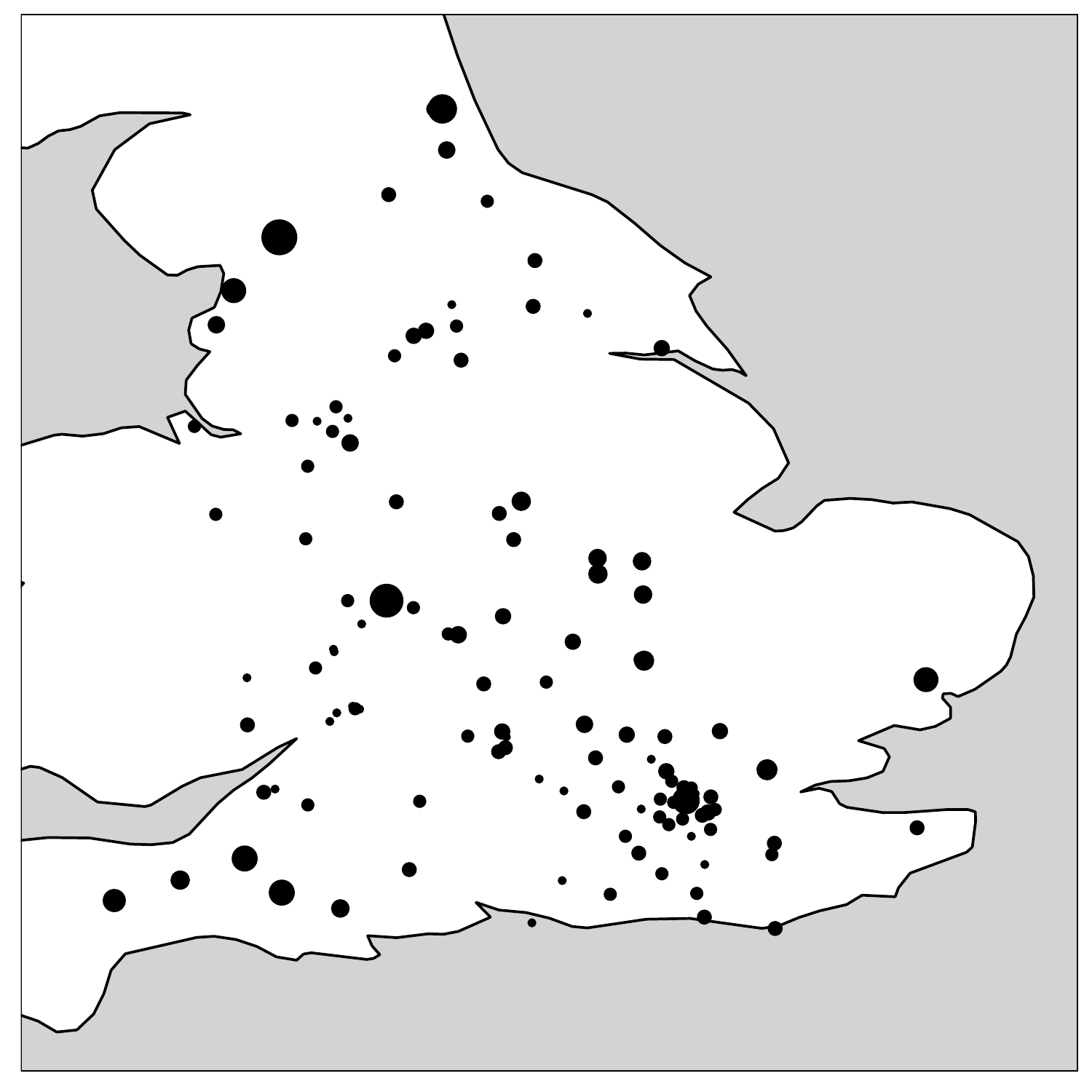}}
	\subfloat{\includegraphics[width=6.5cm,height=6.5cm]{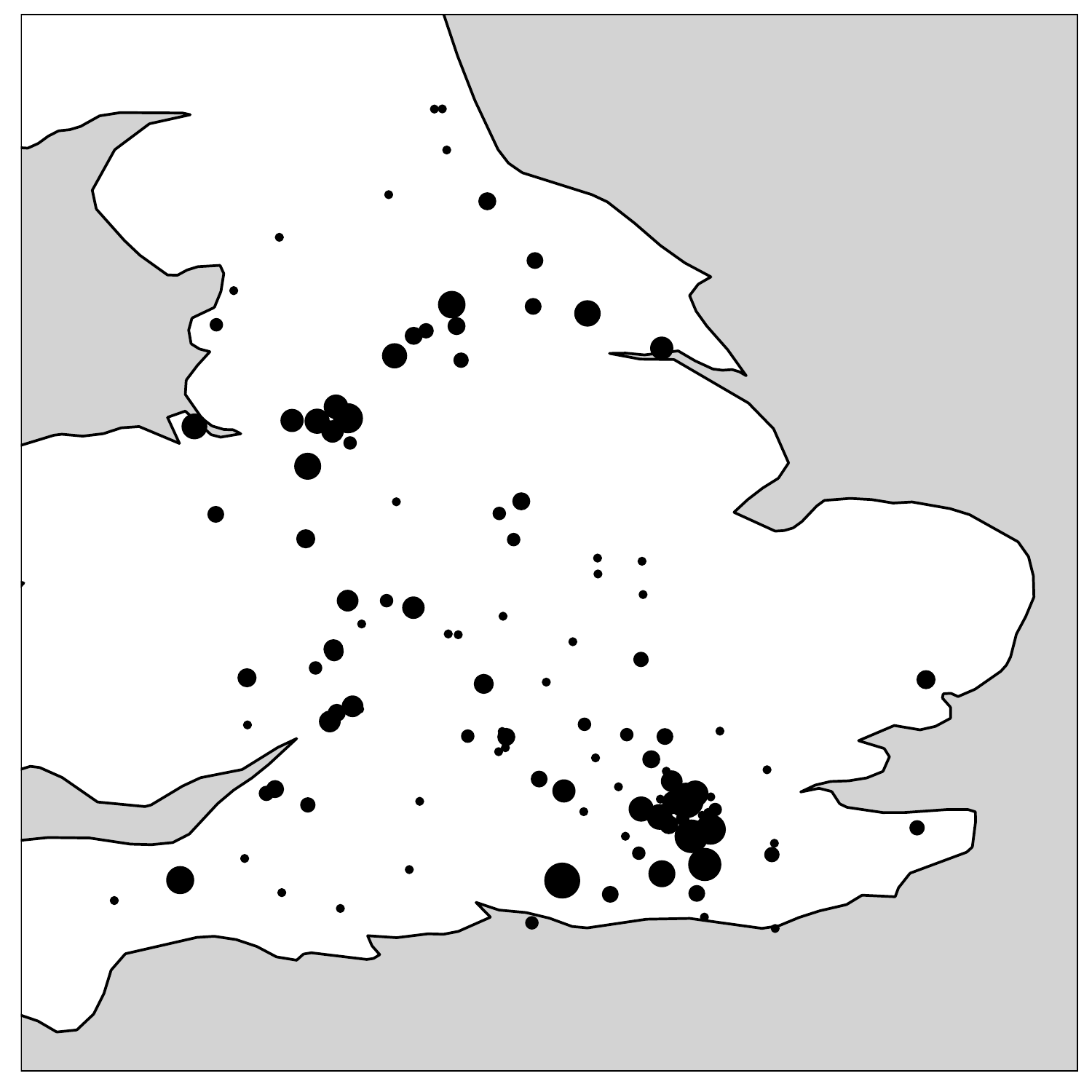}} 
	\caption{Location maps for the teams featuring in the extreme residuals. Size of dot represents the number of times that the team appears in the relevant set of pairs. Left hand chart includes pairs where $d < \hat{d} -4$, and right hand where $d > \hat{d} +4$, where $d$ is the latent space distance and $\hat{d}$ the expected latent space distance based on the linear regression with travel time.}
\label{fig:Maps residuals weighted}
\end{figure}

It might be supposed that the former group is likely to be due to the requirement for geographically extreme teams to travel longer distances in order to complete a sufficient number of matches in the tournament, so that travel time for them has a different level of consideration than for teams with greater geographical proximity to other schools. However Figure \ref{fig:Maps residuals weighted} suggests this does not account for the entire effect. While the most northerly and southerly teams, as well as the single team in East Anglia do stand out in the figure on the left, teams from all over the country are represented and the third highest weighted is King Edward's, Birmingham in the middle of the country, suggesting there is something else at work here. The other group is made up of geographically proximate teams with a large latent space distance. Again it is perhaps to be expected that many of these are in the more densely geographically clustered south east, but some of the most westerly teams also feature strongly. Even in the case of the south east teams, it remains unexplained as to why these particular teams have this greater latent space distance. This will be examined further in Section \ref{sec:LS Covariates} when the influence of the other nodal covariates is investigated.

\begin{figure}[hbp!]
    \centering
    \includegraphics[width=15cm, height = 22.5cm]{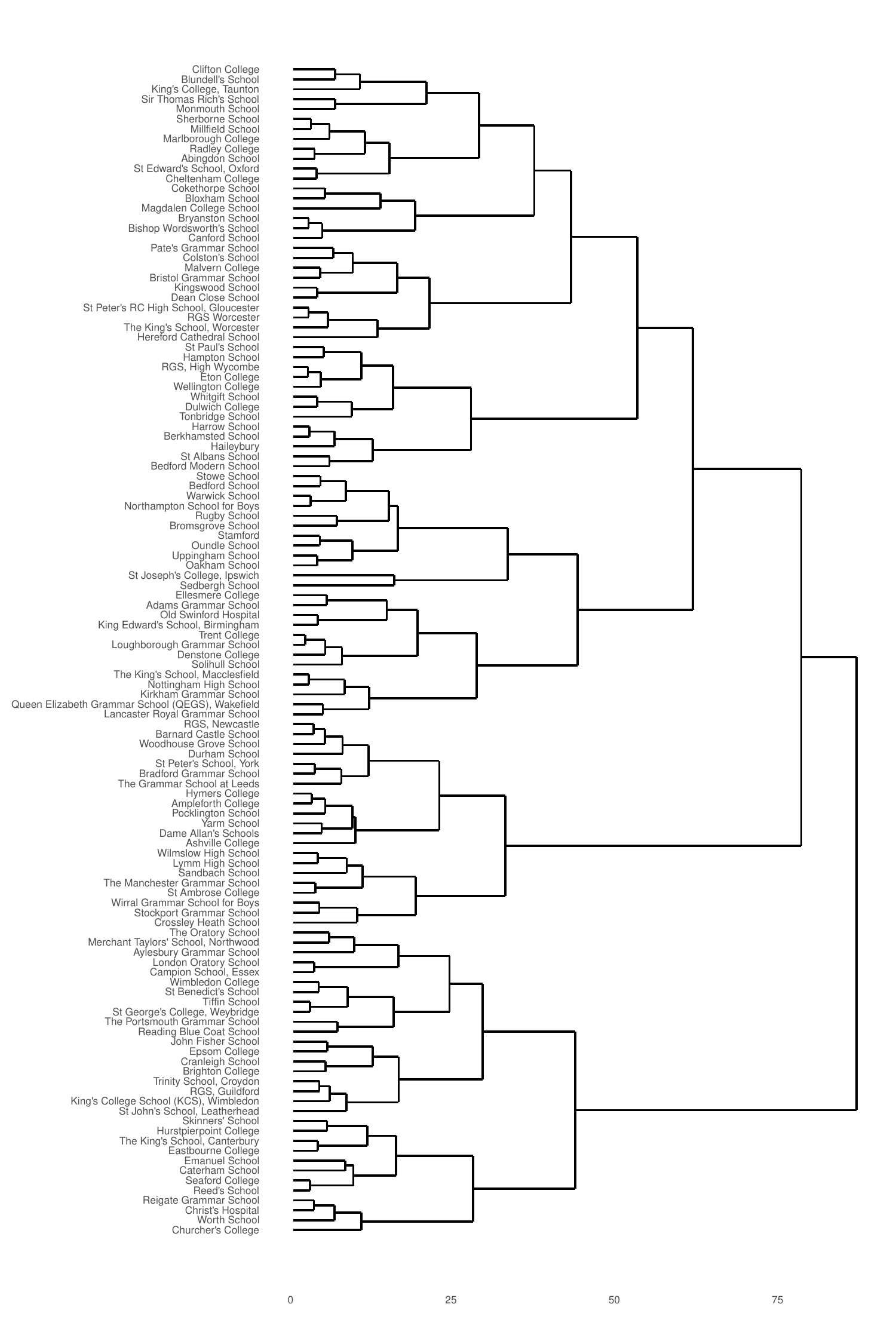}
    \caption{Dendrogram of hierarchical clustering of schools by latent space}
    \label{fig:Dendrogram}
\end{figure}

The geographical implications of model distance may also be considered by investigating community detection in relation to the geographical location of the schools. Numerous methods of clustering could be applied given the latent space distance matrix. As an example, by applying a hierarchical clustering with complete linkage the dendrogram presented in Figure \ref{fig:Dendrogram} is obtained. 

\newgeometry{top=5mm}
\restoregeometry

\begin{figure}
\centering
	\subfloat{\includegraphics[width=0.3\linewidth]{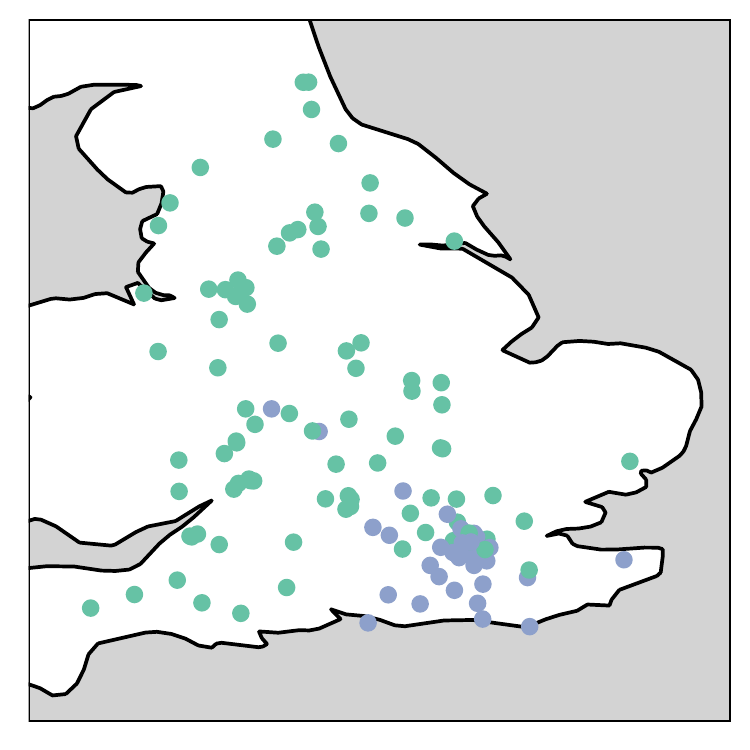}}
	\subfloat{\includegraphics[width=0.3\linewidth]{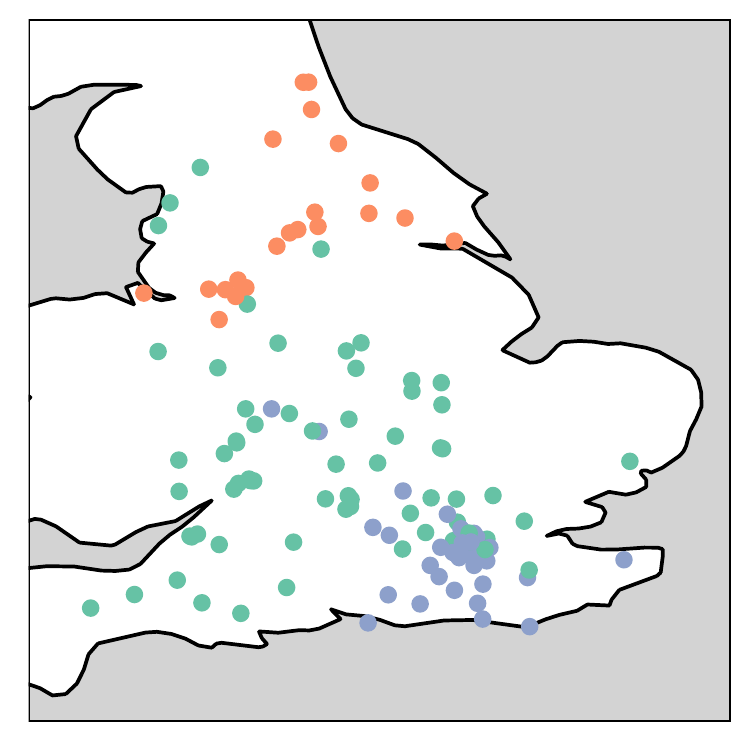}} 
	\subfloat{\includegraphics[width=0.3\linewidth]{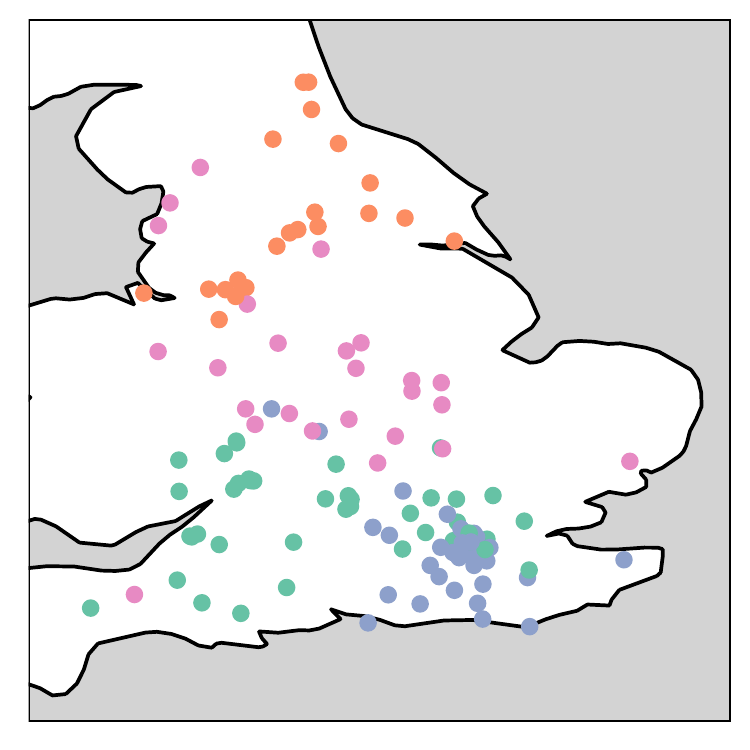}} \\
    \subfloat{\includegraphics[width=0.3\linewidth]{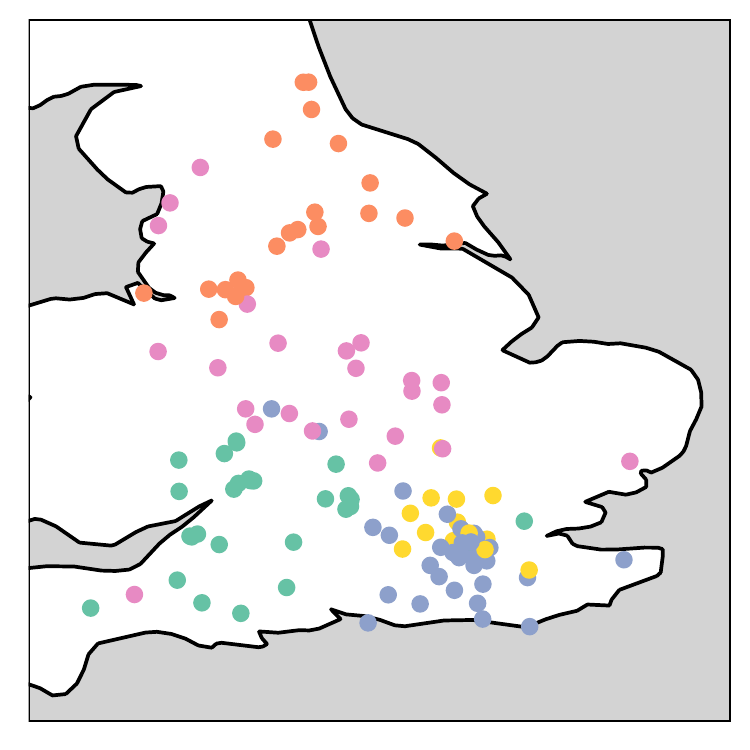}}
	\subfloat{\includegraphics[width=0.3\linewidth]{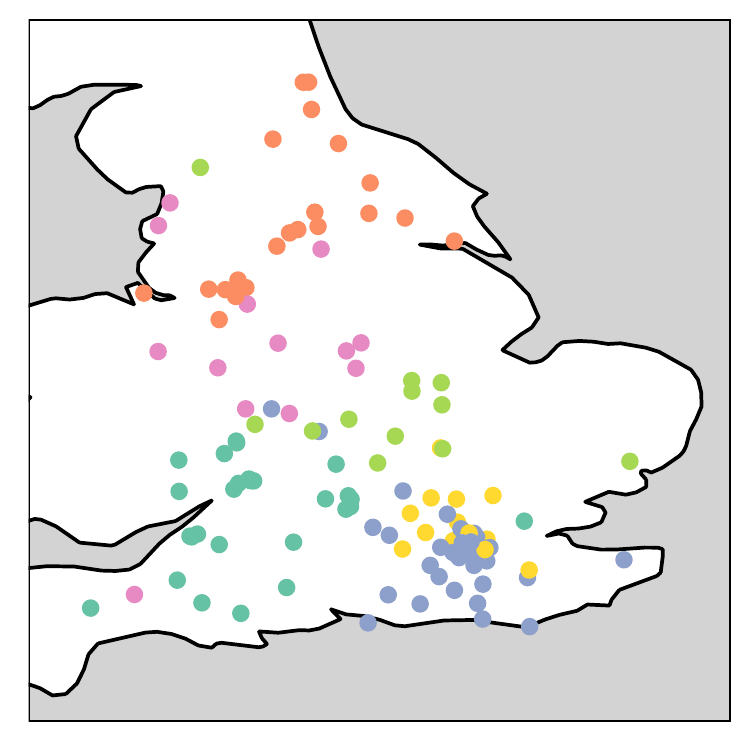}} 
	\subfloat{\includegraphics[width=0.3\linewidth]{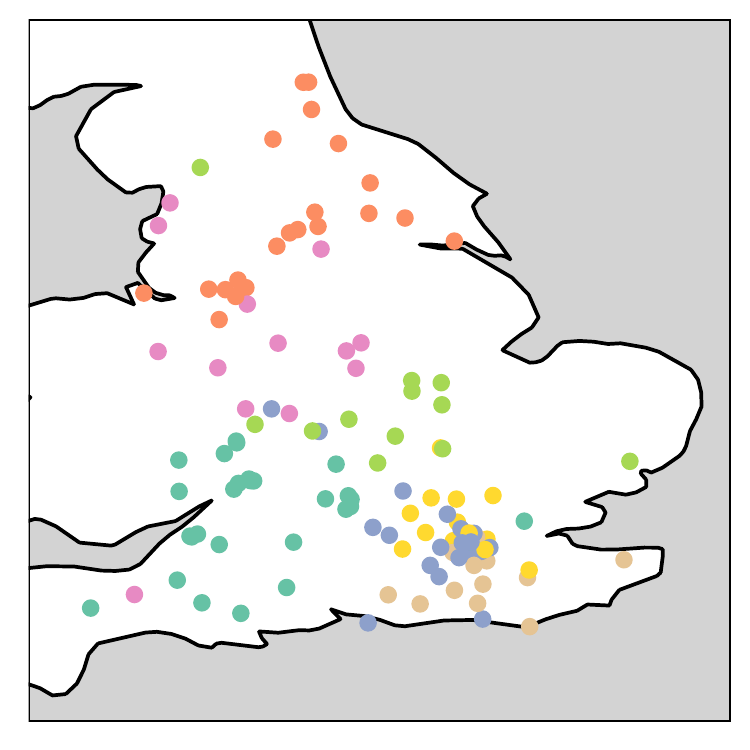}} 
\caption{Communities of size G=2,3,4,5,6,7 based on complete linkage hierarchical clustering of latent space distances}
\label{fig:Maps hierarchical no covariates}
\end{figure}

Inspection of the dendrogram suggests that five communities could be an appropriate partition with a number of communities converging at a distance of around 45. In Figure \ref{fig:Maps hierarchical no covariates} communities are plotted for $G=2,3,4,5,6,7$, where $G$ is the number of groups, in order to show the geographical detection ability with different numbers of groups.

\subsection{Latent Position Cluster Model}\label{sec:LS Simultaneous}
\cite{handcock2007model} provided a model with the extended feature that the latent positions are drawn from a finite mixture of multivariate normal distributions. That is,
\[
\boldsymbol{Z_i} \overset{i.i.d.}{\sim}\sum_{g=1}^{G}\lambda_g\text{MVN}_d(\mu_g,{\sigma_g}^2I_d)\tag{3}\label{eq:3} .
\]
Two methods of estimation were proposed in \citet{handcock2007model}. The first method is a two-stage maximum likelihood procedure, and the second a fully Bayesian estimation using Markov chain Monte Carlo sampling. In this paper, this is fitted based on a burn-in period of 10,000 iterations and a sample run of 1,000,000 iterations of which every fiftieth was sampled, giving a sample size of 20,000.

\begin{figure}
\centering
	\subfloat{\includegraphics[width=0.3\linewidth]{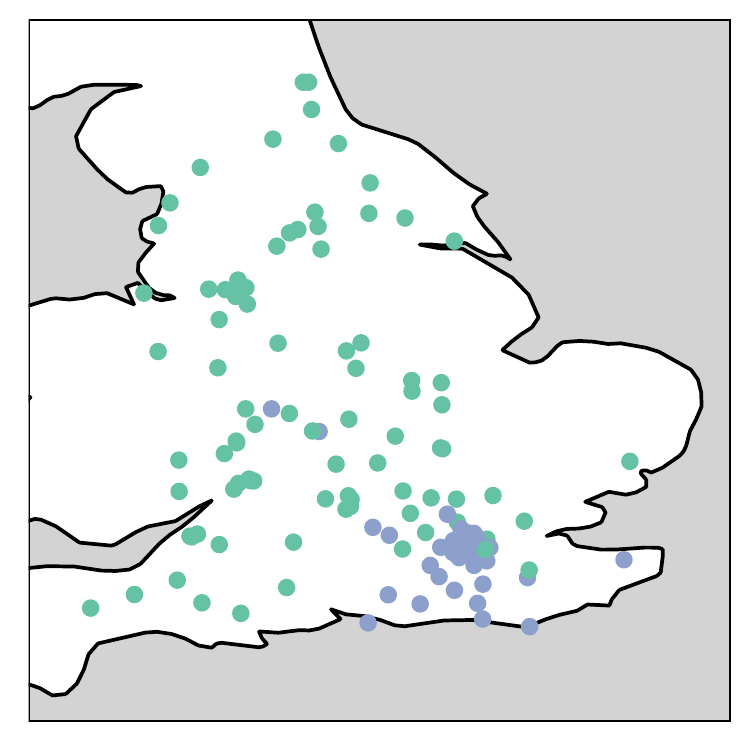}}
	\subfloat{\includegraphics[width=0.3\linewidth]{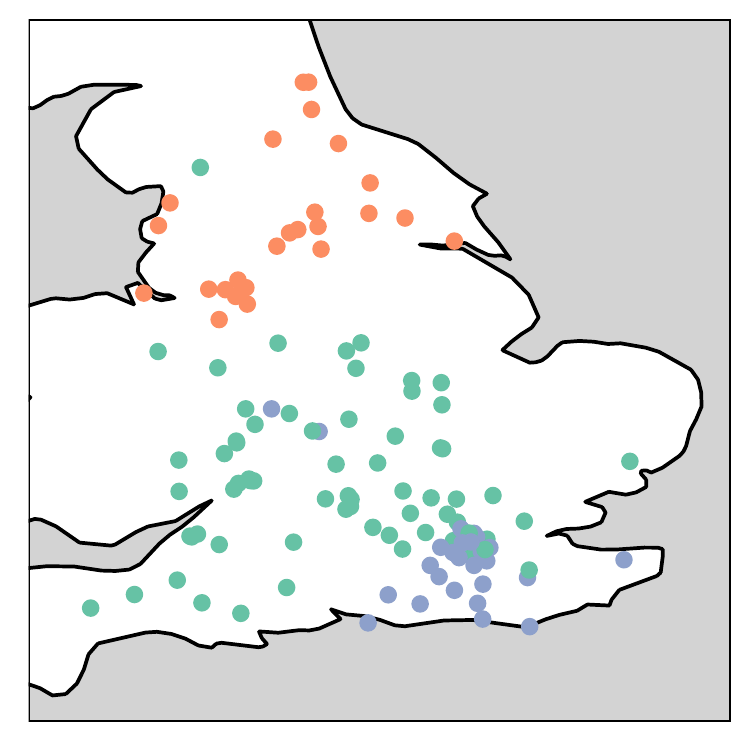}} 
	\subfloat{\includegraphics[width=0.3\linewidth]{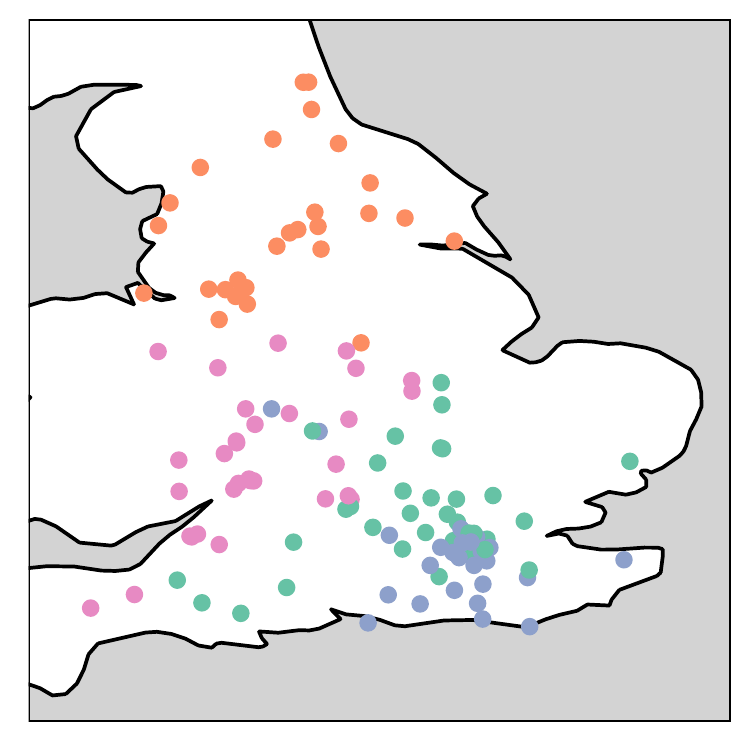}} \\
    \subfloat{\includegraphics[width=0.3\linewidth]{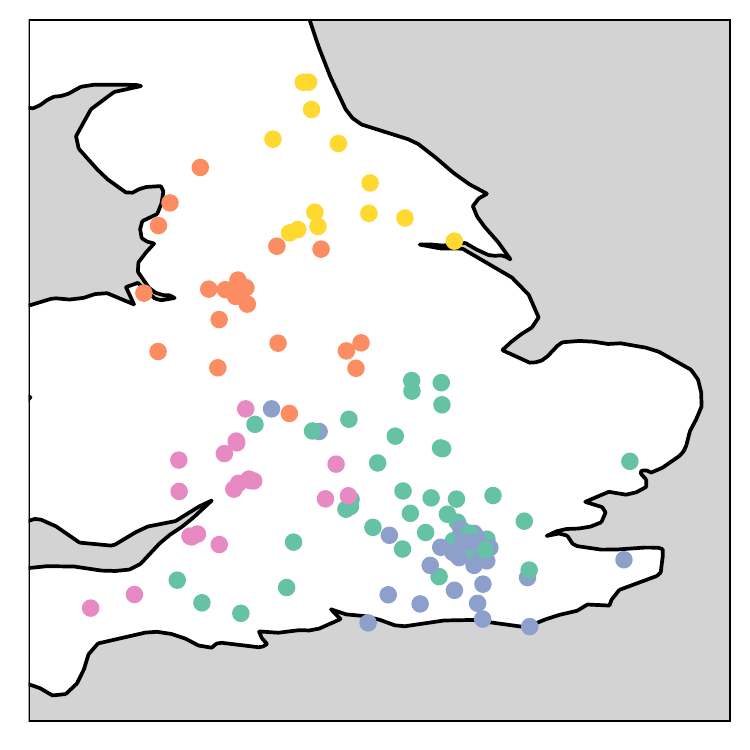}}
	\subfloat{\includegraphics[width=0.3\linewidth]{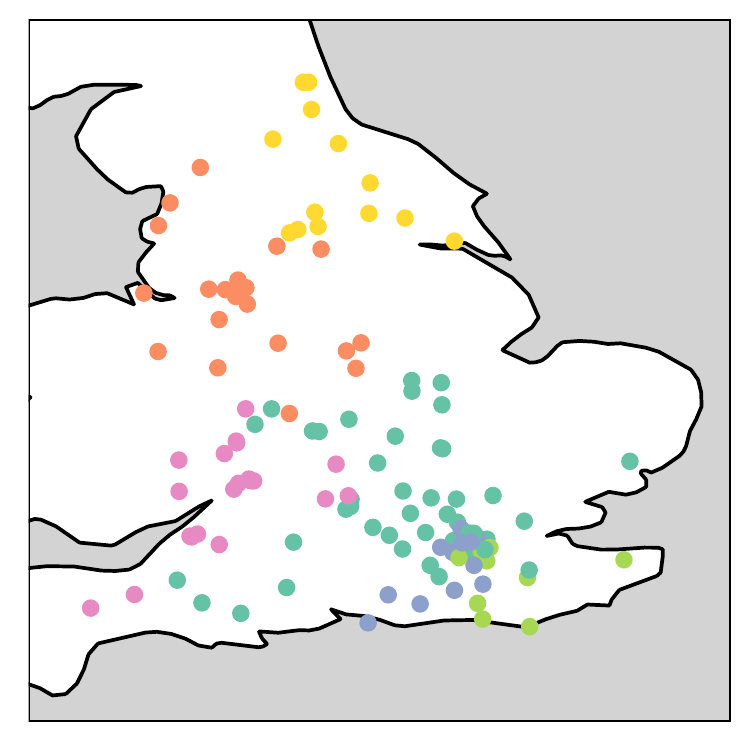}} 
	\subfloat{\includegraphics[width=0.3\linewidth]{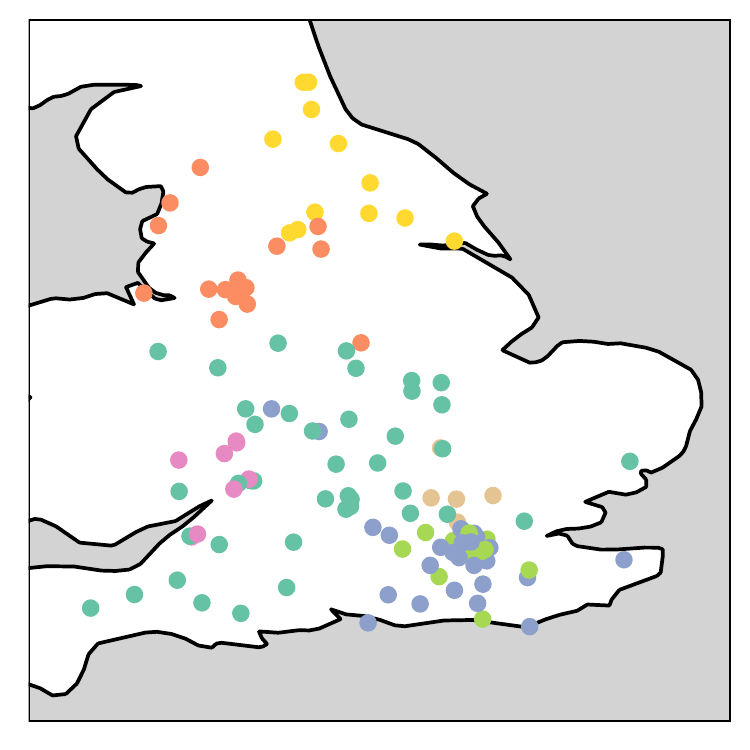}} 
\caption{Communities of size $n=2,3,4,5,6,7$ based on two stage MLE}
\label{fig:Maps MLE no covariates}
\end{figure}

\begin{figure}
\centering
	\subfloat{\includegraphics[width=0.3\linewidth]{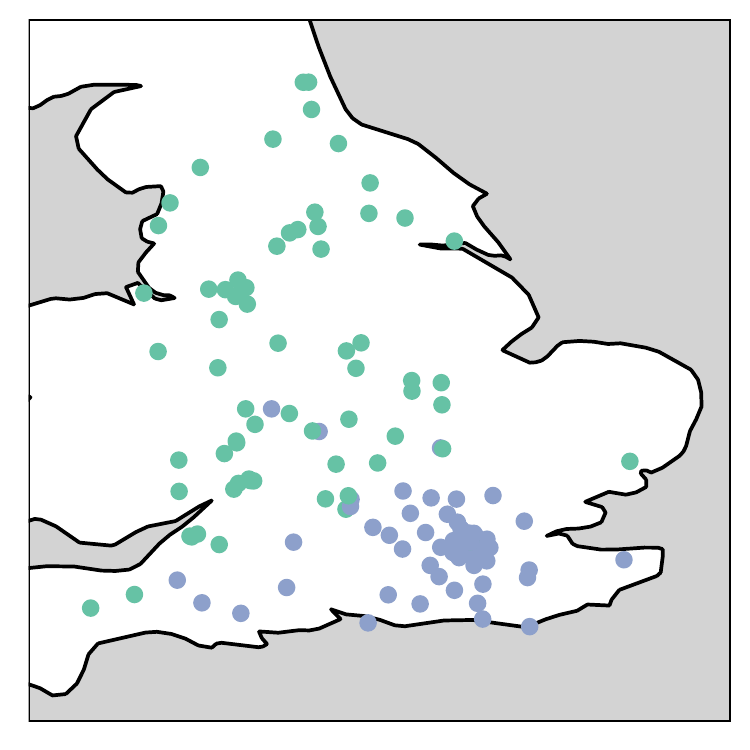}}
	\subfloat{\includegraphics[width=0.3\linewidth]{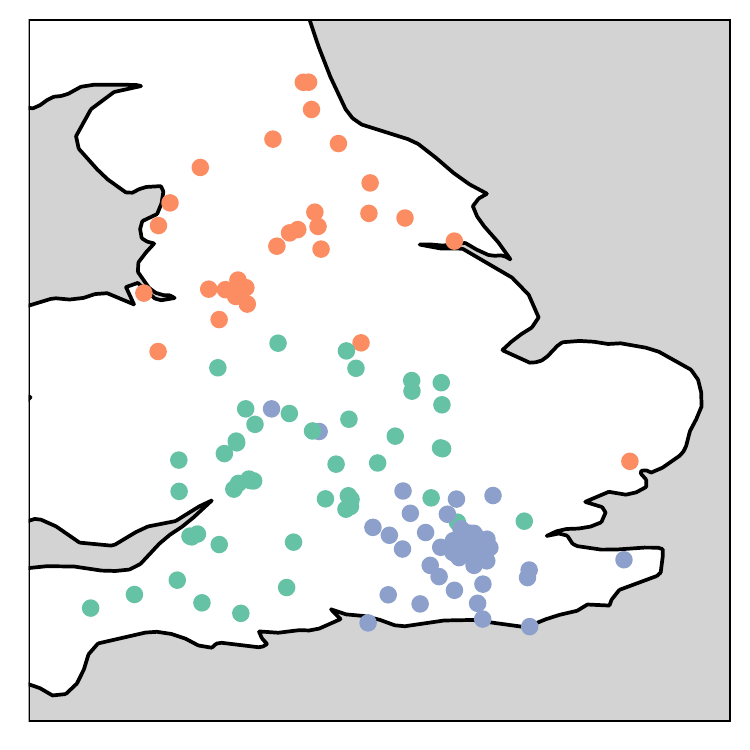}} 
	\subfloat{\includegraphics[width=0.3\linewidth]{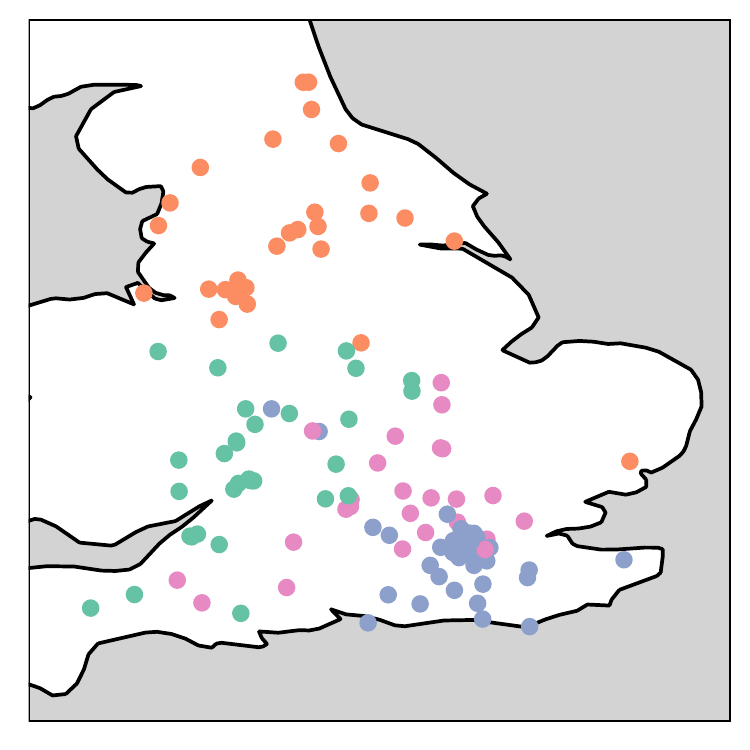}} \\
    \subfloat{\includegraphics[width=0.3\linewidth]{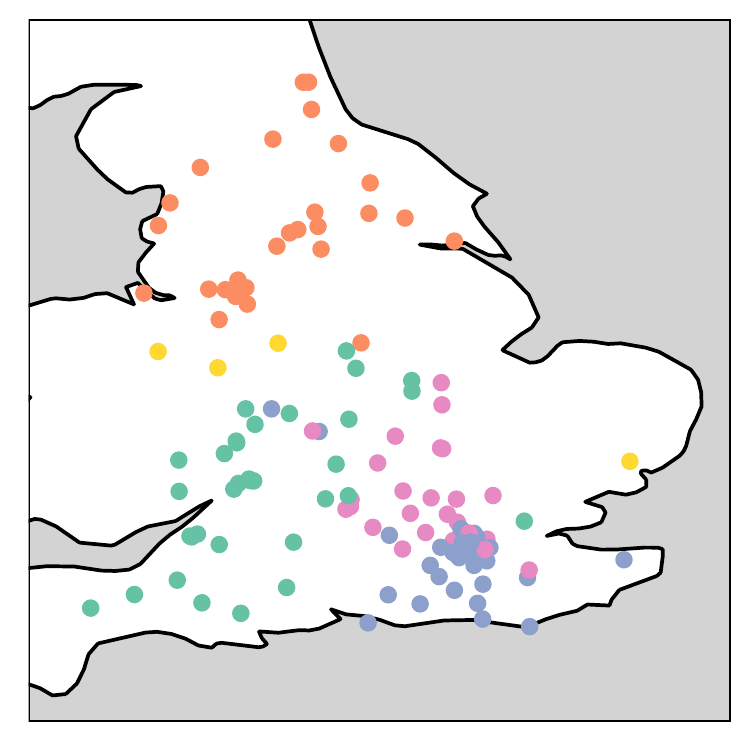}}
	\subfloat{\includegraphics[width=0.3\linewidth]{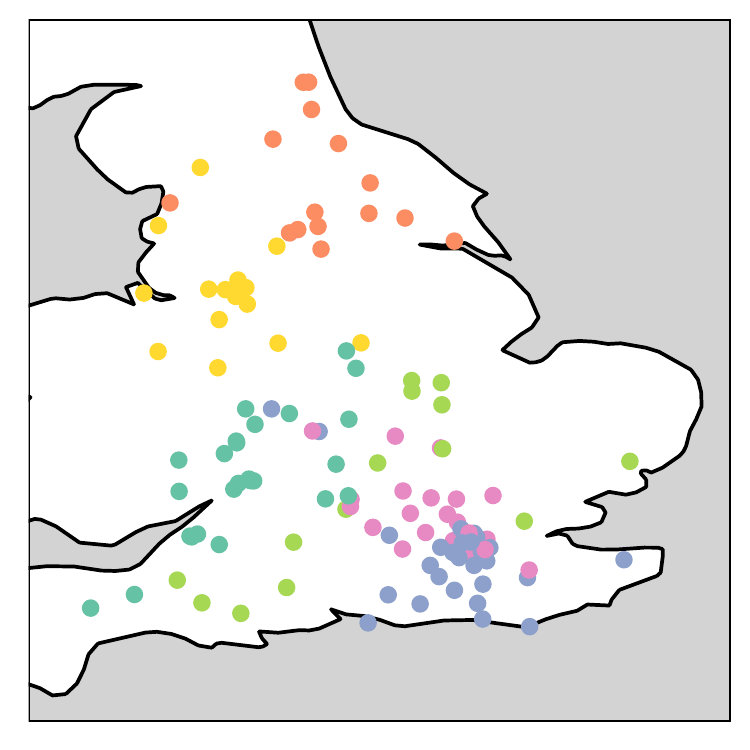}} 
	\subfloat{\includegraphics[width=0.3\linewidth]{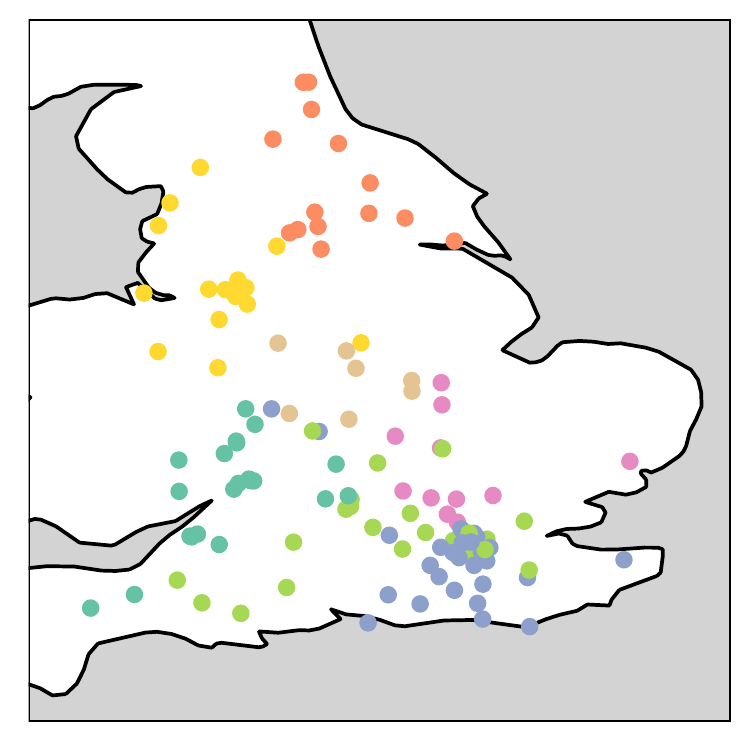}} 
\caption{Communities of size n=2,3,4,5,6,7 based on MCMC parameter estimation}
\label{fig:Maps MCMC no covariates}
\end{figure}

Figures \ref{fig:Maps MLE no covariates} and \ref{fig:Maps MCMC no covariates} present the results from the two methods. In Figures \ref{fig:Maps hierarchical no covariates}, \ref{fig:Maps MLE no covariates}, and \ref{fig:Maps MCMC no covariates}, communities are identified such that the number of schools remaining in the same community as in the previous clustering (as represented by a particular colour) is maximised. The different colours representing different regions in the three Figures is thus a result of different evolutions of the community detection in each case. The fittings using the Gaussian clustering, both based on the two step MLE and the MCMC, appear to show better geographic separation than did the hierarchical clustering. The MCMC clustering arguably shows a better separation in the London area, though there is substantial agreement in the community membership up to $G=5$.  

\begin{figure}
    \centering
    \includegraphics[width=13.5cm, height = 6cm]{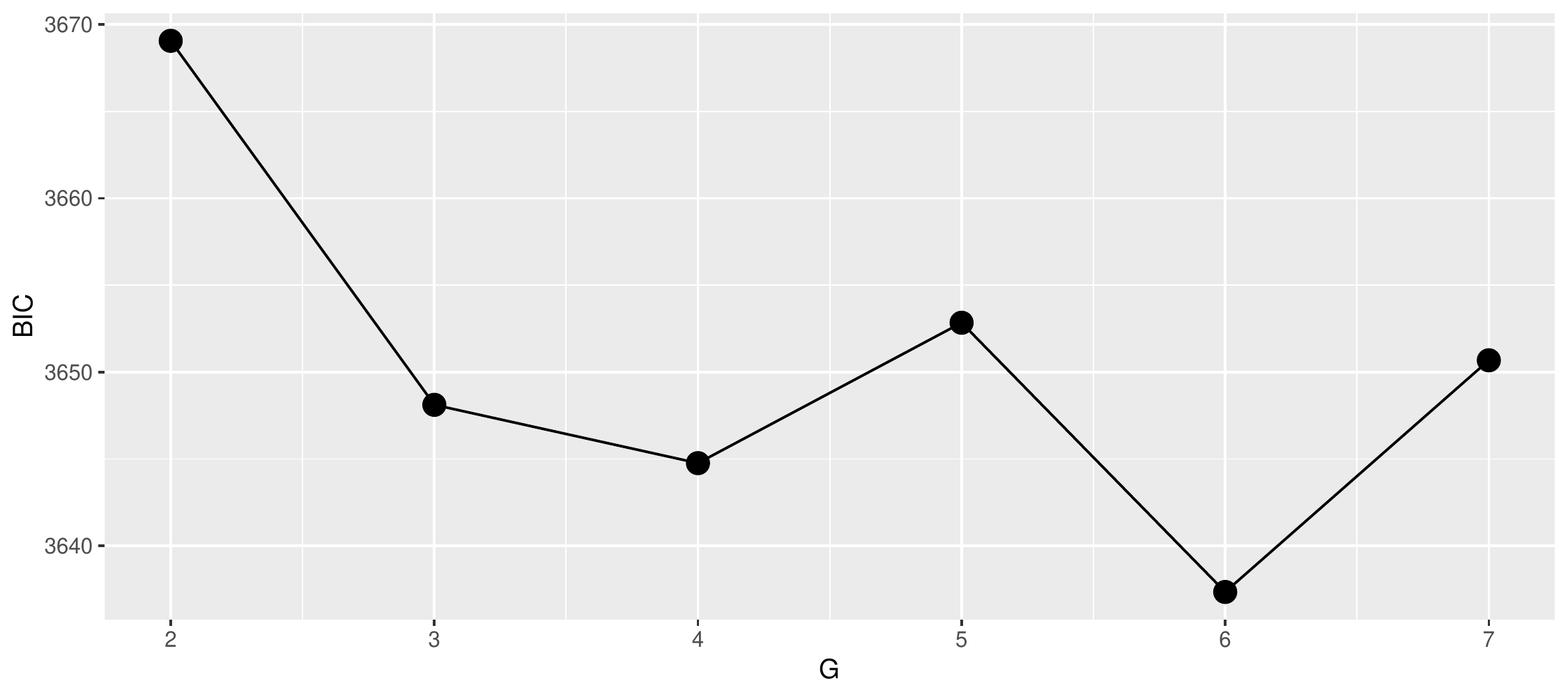}
    \caption{Bayesian Information Criterion for different values of G}
    \label{fig:BIC}
\end{figure}

For the purposes of further investigation it is useful to select a single number of communities, $G$, with which to work. Graphical inspection of Figures \ref{fig:Maps MLE no covariates} and \ref{fig:Maps MCMC no covariates} suggests that between three and five communities may fit best, given the clear geographical separation they evidence. \citet{handcock2007model} suggest the use of a Bayesian Information Criterion (BIC) for the purpose of selecting the number of communities when using the MCMC fitting. Figure \ref{fig:BIC} shows this BIC for each value of $G$. Experimentation with different specifications of the algorithm showed the BIC value to be somewhat unstable. As such this is taken to be indicative rather than definitive and so taking into account both the BIC and the geographical separation $G=4$ is chosen. 

\begin{figure}
\centering
	\subfloat{\includegraphics[width=0.5\linewidth]{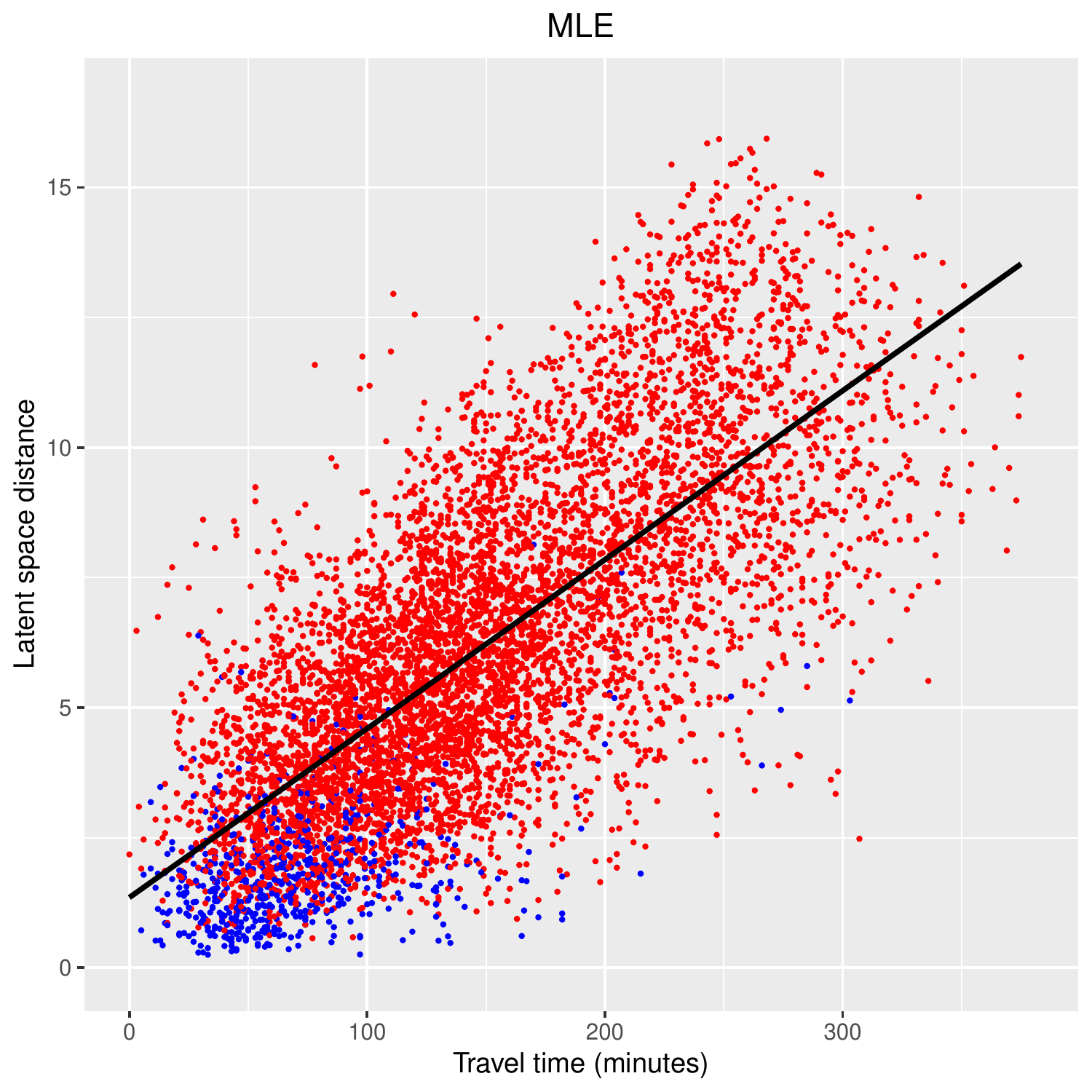}}
	\subfloat{\includegraphics[width=0.5\linewidth]{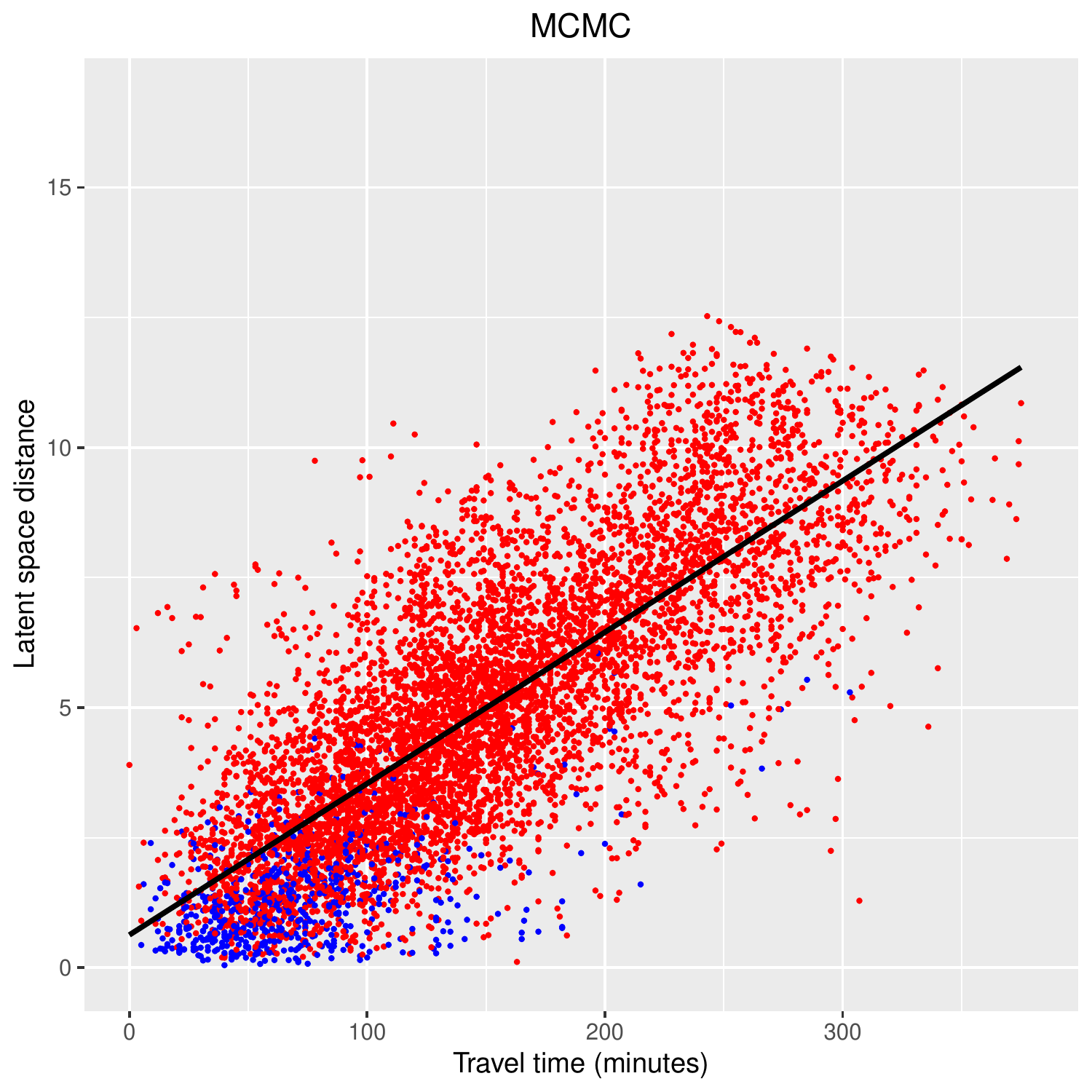}} 
\caption{Scatterplot of latent space distance against travel time for two stage MLE and MCMC fits with $G=4$. School pairs who do not play each other during the three seasons are in red, pairs who play each other at least once are in blue. Linear regression line is shown in black.}
\label{fig: Model distance vs Travel time MCMC}
\end{figure}

In Figure \ref{fig: Model distance vs Travel time MCMC} the travel time in minutes is plotted against the latent space distance. This is shown side by side for the two stage MLE and the MCMC estimation to aid comparison. The most notable feature is that the effect of applying the simultaneous Bayesian estimator obtained from the MCMC fit is to reduce and also compress latent space distances particularly for the large latent space distances. The interquartile range of the residuals, for example, is thus more than 25\% lower when compared to the MLE fit. The Pearson correlation is slightly higher at 78\% compared to 74\% from the MLE fit, and the intercept is closer to zero. This suggests that the assumption of community membership usefully constrains the fit. The community memberships are substantially similar with 108 of the 118 schools belonging to equivalent groups. However based on the seemingly improved latent space fit, the MCMC estimation with $G=4$ is the model selected for further analysis.

\begin{figure}
\centering
	\subfloat{\includegraphics[width=0.5\linewidth]{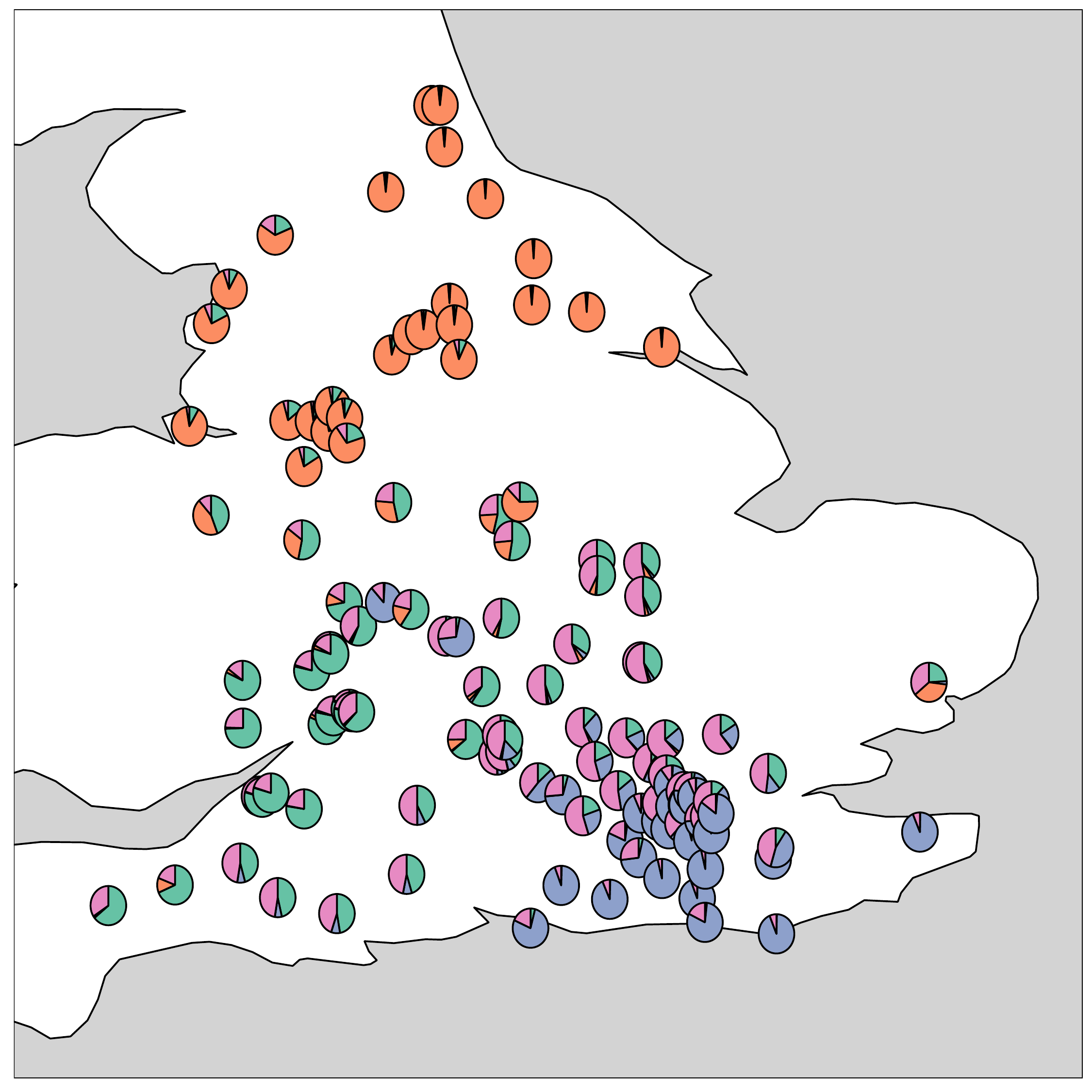}}
\caption{Probability of membership of each community.}
\label{fig: Scatterpie G=4}
\end{figure}

A feature of the mixture model is that it allows us to view the probability of each community membership for each node. Figure \ref{fig: Scatterpie G=4} provides a graphical interpretation of these probabilistic community memberships. Perhaps the most notable feature is the difference in delineation of the communities. What might be referred to as the northern community is most clearly delineated, with all but two members having greater than 75\% probability of being part of that community. On the other hand, the two central communities are substantially more indeterminate with the southernmost (that represented in the chart in pink) having no member with a greater than 75\% probability of being part of the community. The schools in and around London show perhaps more delineation than one might expect given the short distances. Due to the geographical proximity of a number of schools, some are obscured in the chart. This does not alter the interpretation when the distributions of these schools are known. It could also be noted that the most geographically isolated school, St Joseph's Ipswich, has a substantial probability of being in three of the four different groups.

\begin{figure}
    \centering
    \includegraphics[width=12cm, height = 12cm]{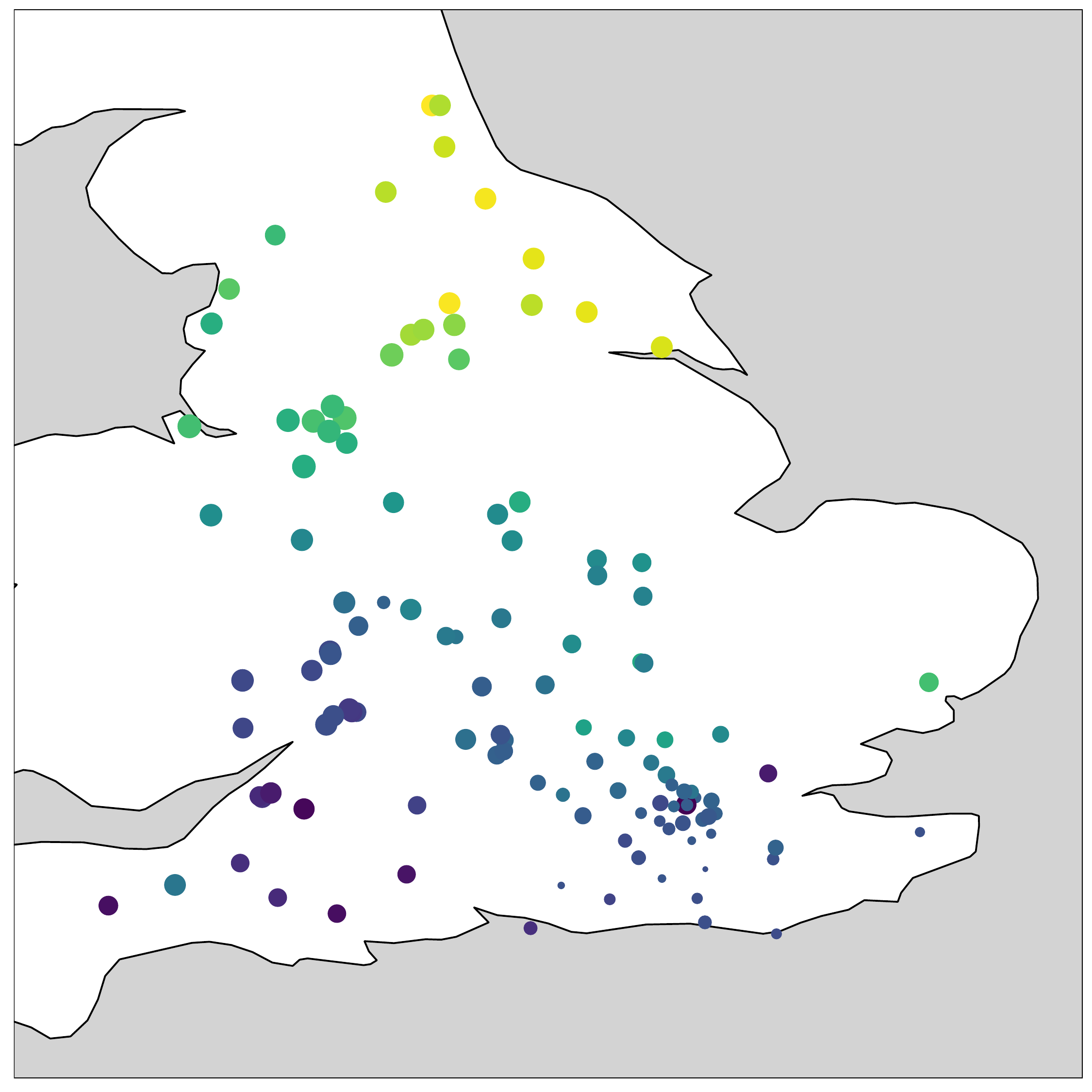}
    \caption{Map of schools with colour and size representing first and second latent space coordinate respectively for each school, based on MCMC estimation of latent position cluster model with $G=4$}
    \label{fig: LS Map}
\end{figure}

An alternative way of considering the link between the latent and geographical spaces is presented in Figure \ref{fig: LS Map}. This suggests that the two latent space dimensions are picking up something like a polar coordinate system with a radial coordinate eminating from a centre in the south-east (as seen on the size scale), and an angular coordinate going from the south to the north east (as seen in the colour scale), an observation which is consistent with the probabilistic clustering, as well as the evolution of the community detection seen in Figures \ref{fig:Maps MLE no covariates} and \ref{fig:Maps MCMC no covariates}.

\section{A consideration of covariates}\label{sec:LS Covariates}

\subsection{Relative Importance} \label{sec:relative importance}
The analysis so far has shown a clear link between the travel time and the propensity for the existence of fixtures, but this is to be expected and it is desirable to consider how it compares to the other covariates of interest. \citet{gromping2015variable} provides a thorough overview of various measures of relative importance, and the factors to consider in selecting one. These are presented in the context of linear regression. Some of these measures may be extended to wider families of models, and in particular, \citet{thomas2008measuring} proposed relative importance measures for logistic regression that are consistent with some of the features of the measure axiomatically justified by \citet{pratt1987dividing} for linear regression.

In the present case, the unreliability and the computational expense of the coefficient estimates obtained when the model is fitted with all covariates simultaneously are considerable hindrances to applying relative importance measures directly to the model. This is especially so since a number of the measures require at least $p!$ computations, where $p$ is the number of regressor variables, in order to average over the permutations of regressor variables. One alternative is instead to use the pairwise latent space distances estimated by the selected model. A linear regression of these distances against the nodal covariates can then be considered. Doing so provides an estimate for the relative importance of the covariates in the determination of pairwise latent space distance. One reason to be cautious of this approach is that it may be viewed as a measure of relative importance of the covariates only in so much as latent space distance is a good measure of the propensity for a fixture to exist. However if the conclusions drawn from the analysis prove to be corroborated by other approaches then we may have more confidence about the meaningfulness of the results.

Two specific measures are considered here. Both have the desirable and intuitive properties of providing: independence from the order of the regressors in the model; scale invariance; a decomposition of the model variance ($R^2$); and a proper decomposition of the model variance for any orthogonal regressor subgroups. The first (which we will refer to as `Pratt') is the one advocated by \cite{pratt1987dividing}, which provides an axiomatic justification for using $b_k\rho_k$ to assess relative importance, where $b_k$ is the standardised coefficient for the $k$th regressor variable, $x_k$, and $\rho_k$ is the correlation of $x_k$ with the independent variable $y$, in this case the latent space distance. The principal objection to this approach is that it can produce a negative measure, which is not clearly interpretable. The second (LMG, after the original authors) is the method originally due to \cite{lindeman1980introduction}, which was also influentially and independently advocated by \cite{kruskal1987relative}. For each regressor, this takes the marginal increase of explained variance from the addition of the regressor to the model, and takes the mean of these values over all regressor order permutations of the model. Taking the notation of \cite{gromping2015variable}, denote the explained variance and the sequential additional variance respectively as
\begin{align*}
    \text{evar}(S) &=\text{var}(y)-\text{var}(y \mid x_j;j \in S),  \\
    \text{svar}(M) &=\text{evar}(M \cup S) - \text{evar}(S).
\end{align*}
Then we can define, without loss of generality, the measure for the first regressor as
\[
\text{LMG}(1) = \frac{1}{p!}\sum_\pi \text{svar}(\{1\} \mid S_1(\pi)),
\]
where $\pi$ are the regressor permutations and $S_1(\pi)$ the set of regressors preceding regressor 1 in permutation $\pi$. 
The model has an $R^2$ of 62\%, and the proportion of that coming from each covariate is shown in Table \ref{tab:LSDistance Relative Importance}.

\begin{table}[htbp!]
\centering
\begin{tabular}{lcc}
\hline
Variable        & Pratt     & LMG \\
\hline
Travel time     & 98.09\%   & 97.77\% \\
Percent Boarder & 0.88\%    & 0.88\%  \\
Fees            & 0.72\%    & 0.63\%  \\
Percent Boys    & 0.09\%    & 0.08\%  \\
Boys            & 0.08\%    & 0.12\%  \\
School type     & 0.05\%    & 0.30\%  \\
Rating          & 0.04\%    & 0.14\%  \\
Term type       & 0.02\%    & 0.02\%  \\
Founded         & 0.01\%    & 0.04\%  \\
\hline
\end{tabular}
\caption{Proportion of explained variance ($R^2$) attributable to each covariate in linear regression of pairwise latent space distance against covariates. Ordered in descending order of relative importance under the Pratt measure.}
\label{tab:LSDistance Relative Importance}
\end{table}

It is perhaps unsurprising based on the previous analysis that travel time is dominant, but the degree to which this is the case is nevertheless notable. The other covariates have negligible relative importance in comparison to travel time, but relatively they suggest that fees and the proportion of boarders may have greater influence. The two measures of relative importance assessed here, Pratt and LMG, are very substantially consistent, with Pratt giving slightly higher values for travel time and fees and lower for school type and rating. 

\subsection{Covariate inclusion} \label{sec:covariate inclusion}
An alternative means to investigate the influence of the different covariates is by considering models with each covariate included individually,
\[
\text{logit}(\mu_{ij}) = \beta_0 + x_{ijk}\beta_{k} - d(\boldsymbol{Z_i},\boldsymbol{Z_j}), \quad \quad (k=1,\dots p),\tag{4}\label{eq:4}
\]
where the $k$ represent the nodal covariates. Table \ref{tab:MCMC5 Covariate individual estimates} presents the parameter estimates based on the posterior mean of the coefficient, as well as a statistic representing $q = 2 \times \text{min}(P(\beta_k>0),P(\beta_k<0))$, where $\beta_k$ is the coefficient under consideration, and the Bayesian Information Criterion (BIC) for each model, which may be compared with a BIC of 3647 for the model with no covariates included. Comparing individually in this way thus allows us to consider the importance of a particular covariate based on the change in BIC, and the $q$-value statistic. 


\begin{table}[htbp!]
\centering
\begin{tabular}{p{0.2\textwidth}>{\centering}p{0.2\textwidth}>{\centering}p{0.2\textwidth}>{\centering\arraybackslash}p{0.2\textwidth}}
\hline
& $\hat{\beta_k}$ & $q$ & BIC\\
\hline
Travel Time            & $-$1.523           & $<10^{-15}$    & 3316\\
Fees            & $-$0.368           & $<10^{-15}$    & 3618\\
Percent boarder & $-$0.860           & $<10^{-15}$   &  3630\\
Term type       & $-$0.309            & 0.0016 & 3642      \\
6th Form boys   & $-$0.199           & 0.0039 & 3642  \\
School type     & $-$0.370            & 0.0017 & 3643      \\
Founded         & $-$0.038           & 0.0473 & 3646                 \\
Percent boys    & $-$0.756           & 0.0070 & 3649                 \\
Rating          & $-$0.057           & 0.3239  & 3651       \\          
\hline                 
\end{tabular}
\caption{Coefficient, $q$-value and BIC when model fitted with individual additional covariates. Ordered in increasing BIC.}
\label{tab:MCMC5 Covariate individual estimates}
\end{table}

The directions of the coefficient estimates are as one would expect, so that a smaller covariate distance implies a higher probability of a fixture for all covariates. The BIC estimates are consistent with what was seen in the relative importance analysis in highlighting travel time as the dominant covariate. In that fees and the proportion of boarders are the only others that have a notably lower BIC and also $q$-value, these are also somewhat consistent with the previous results. With the exception of rating, the $q$-values are all low,  indicating that it is likely that these factors are influencing the propensity for a fixture to exist. This is perhaps surprising in the case of year of foundation, though the $q$-value there is less conclusive.

Alternatively we may choose to fit the model with all covariates, as in equation \ref{eq:5}.
\[
\text{logit}(\mu_{ij}) = \beta_0 + \sum^p_{k=1}x_{ijk}\beta_{k} - d(\boldsymbol{Z_i},\boldsymbol{Z_j}),\tag{5}\label{eq:5}
\]
The results are presented in Table \ref{tab:MCMC4 Covariate together estimates}. The computation of this shows greater sensitivity to the specification of the algorithm and even the order in which covariates are presented, so we might be cautious about the results. They do however show broad agreement with previous analyses, and where there are discrepancies between the coefficient estimates when taken together and individually these may reasonably be thought of as being as a result of the confounding of covariates. For example, the correlation between the percentage of boarders and fees provides an explanation for the smaller absolute impact from boarding when the covariates are considered together. 

In this context however it is perhaps more intuitive to interpret the impact of the covariates individually as in the models represented by equation \ref{eq:4}, in the sense that knowing nothing else about the schools in each case one would expect the odds of a fixture to decrease by 78\% for every additional hour of travel between schools, by 31\% for every £10,000 per annum difference in fees, and by 58\% for a fully boarding school (100\% boarders) playing a fully day school (0\% boarders) as compared to a match between two schools with the same proportion of boarders.

\begin{table}[htbp!]
\centering
\begin{tabular}{p{0.2\textwidth}>{\centering}p{0.2\textwidth}>{\centering\arraybackslash}p{0.2\textwidth}}
\hline
& $\hat{\beta_k}$ & $q$ \\
\hline
Travel Time         & $-$1.552          & $<10^{-15}$      \\
Fees                & $-$0.363          & $<10^{-15}$      \\
Percent boarder     & $-$0.317          & $<10^{-15}$      \\
Term type           & $-$0.283          & $<10^{-15}$      \\
6th Form boys       & $-$0.252          & $<10^{-15}$      \\
Percent boys        & $-$0.734          & $<10^{-15}$      \\
Founded             & $-$0.042          & 0.0478           \\
School type         & $+$0.154            & 0.1541           \\
Rating              & $-$0.002          & 0.9613           \\          
\hline                 
\end{tabular}
\caption{Coefficient $q$-values when model fitted with all covariates. Ordered in increasing $q$-value.}
\label{tab:MCMC4 Covariate together estimates}
\end{table}

\subsection{Graphical inspection} \label{sec:graphical inspection}

Another method of investigation that the model allows is by inspection of the latent space graphically. Given the dominance of travel time, it is useful to control for it in considering the other covariates. Thus the following model is considered with G=4 based on BIC values.
\[
\text{logit}(\mu_{ij}) = \beta_0 +  x_{ij\text{Travel Time}}\beta_{\text{Travel Time}} - d(\boldsymbol{Z_i},\boldsymbol{Z_j}),\tag{6}\label{eq:6} 
\]
The colours are used to show the fees and proportion of boarders for each school. As was noted in Section \ref{sec:Data} these covariates are strongly linked and this is evident in the plots. It can be seen that while schools with lower fees and low proportions of boarders are distributed quite evenly throughout the latent space, those with higher fees and a high proportion of boarders are to be found disproportionately often in a cluster at the bottom of the latent space. This effect is perhaps clearer for the proportion of boarders. This suggests that there is a small community of schools with a higher proportion of boarders and higher fees, who are more likely to play each other, and that it is this that was driving the greater relative importance of these covariates noted in sections \ref{sec:relative importance} and \ref{sec:covariate inclusion}. 
\begin{figure}
\centering
	\subfloat{\includegraphics[width=0.5\linewidth]{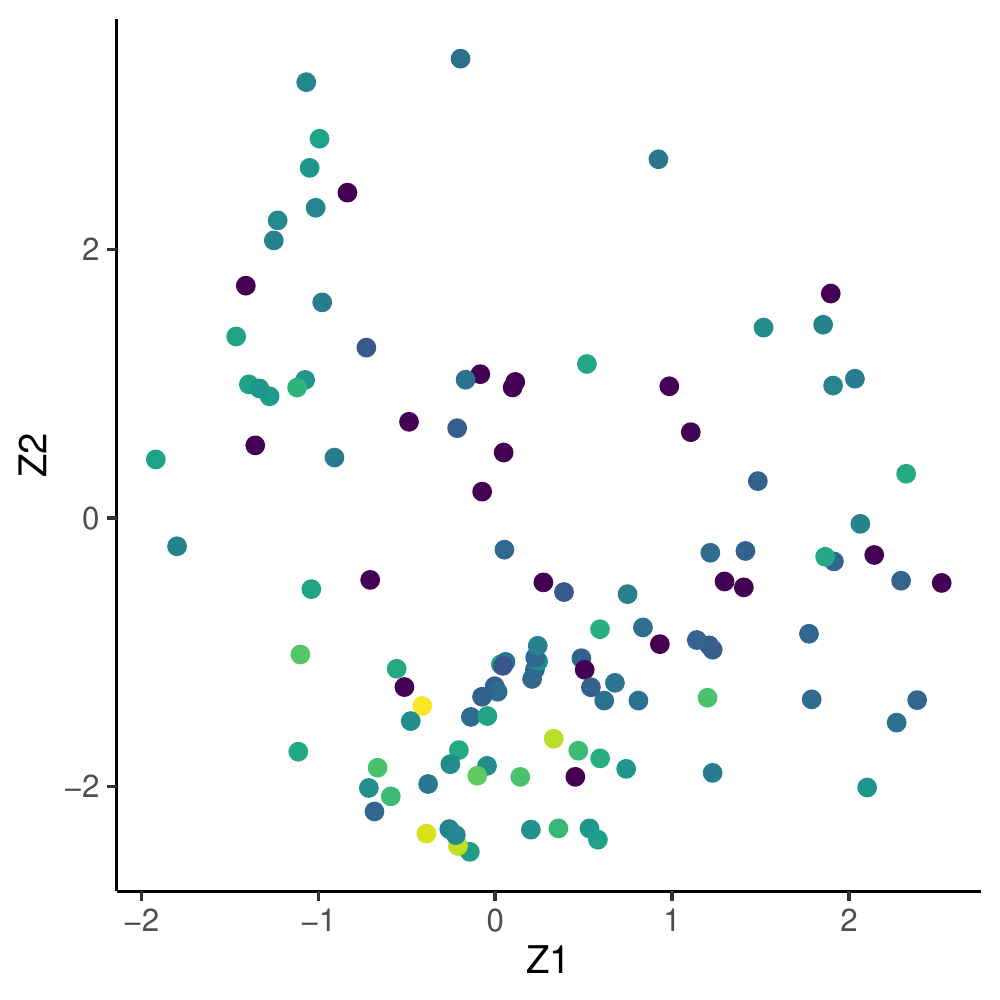}}
	\subfloat{\includegraphics[width=0.5\linewidth]{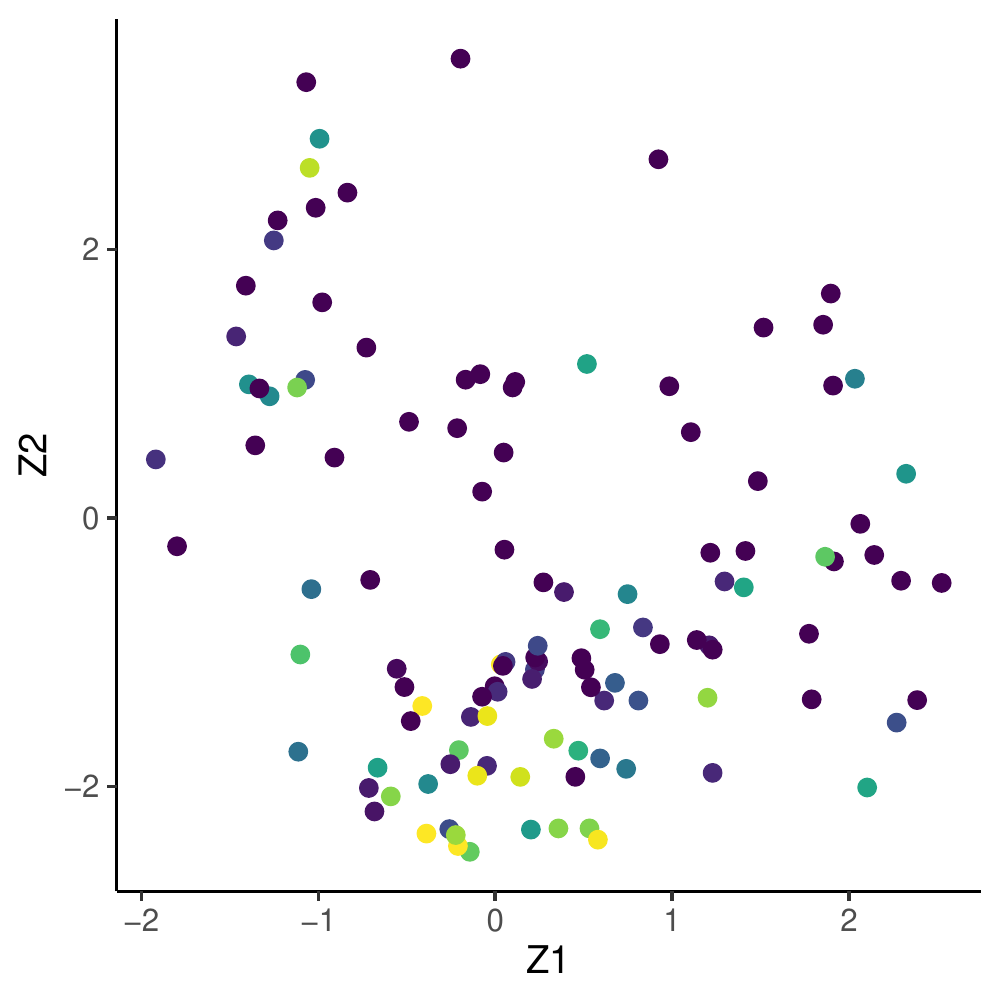}} 
\caption{Latent space plots based on model with Travel Time controlled for and with G=4. Colour scale representing Fees on left hand side, and \% Boarders on right hand side}
\label{fig: LS plots}
\end{figure}

\section{Concluding Remarks} \label{sec:Concluding Remarks}
The latent space models demonstrated a number of desirable features in the context of addressing the question of the nature of institutional ties. They showed a good ability to detect geographical communities in the data set, and the spatial nature of the output provided a natural means of interpreting the result in terms of pairwise distances. It was possible to employ a variety of methods in order to analyse the relative importance of the nodal covariates in the propensity for a fixture to exist --- by using the methods of \citet{pratt1987dividing} and \citet{lindeman1980introduction} to provide quantitative estimates based on latent space distance; by inspection of $q=2 \times \text{min}(P(\beta_k>0),P(\beta_k<0))$ and BIC on fitting models with covariates included; and by using graphical assessment based on latent space plots as in Figure \ref{fig: LS plots}. These produced a consistent interpretation. The latent position cluster model of \cite{handcock2007model} was found to usefully constrain the latent space model such that it appeared to provide a better fit. It also usefully facilitated a means of assessing the strength of community attribution at a school level by producing the probabilities of each school being a member of each community as seen in Figure \ref{fig: Scatterpie G=4}. It is notable that, while a relatively high number of iterations was chosen to be run based on the diagnostics, most of these conclusions were substantially unchanged when running off the default settings of burn-in of 10,000, with 40,000 sampling iterations of which every tenth was sampled giving a sample size of 4,000. The BIC values and coefficient ranges were however notably more affected by changes in other algorithm specifications when run on a lower number of simulations. The latent space models thus performed well in providing quantitative and qualitative insights into the nature of communities and the influential elements of edge formation in this network. 

Perhaps unsurprisingly travel time was found to be the dominant factor in contributing to the prevalence of fixtures. The degree of this dominance is notable however, representing 98\% of $R^2$ based on the two relative importance measures employed. The proportion of boarders and fees were identified as the next most important factors, with the dependency driven by a community of schools with the highest proportion of boarders and the highest fees. It should be noted however that the data represent a self-selecting group of the top rugby-playing schools. It seems not unreasonable to expect that were a full set of data available then the impact of some of the other covariates would take on a greater importance. For example, over a wider data set with more state schools and a greater range of abilities, perhaps fees, school type or rating would take on greater relative importance. However for the question at hand, it is still notable that rather than clustering seeming to correlate with, for example, rating, as meritocrats might have liked to believe, or with school type as others might have suspected, it was a school's status as a majority-boarding school, with a high level of fees, that produced a detectable clustering effect. This might be seen as consistent at an institution level with a traditional understanding of the old boy network.




\bibliography{bibliography.bib}

\end{document}